\pdfoutput=1

\documentclass[11pt,twoside,a4paper,cmspaper,final,collab]{cms-tdr}
\def\svnVersion{146151}\def\cmsCernNo{2012-127}\def\cmsCernDate{\today}\def\cmsMessage{Submitted to the Journal of High Energy Physics}

\begin{document}\cmsNoteHeader{SUS-11-020}

\hyphenation{had-ron-i-za-tion}
\hyphenation{cal-or-i-me-ter}
\hyphenation{de-vices}

\RCS$Revision: 122473 $
\RCS$HeadURL: svn+ssh://alverson@svn.cern.ch/reps/tdr2/papers/SUS-11-020/trunk/SUS-11-020.tex $
\RCS$Id: SUS-11-020.tex 122473 2012-05-17 13:55:20Z alverson $
\newlength\cmsFigWidth
\ifthenelse{\boolean{cms@external}}{\setlength\cmsFigWidth{0.85\columnwidth}}{\setlength\cmsFigWidth{0.4\textwidth}}
\ifthenelse{\boolean{cms@external}}{\providecommand{\cmsLeft}{top}}{\providecommand{\cmsLeft}{left}}
\ifthenelse{\boolean{cms@external}}{\providecommand{\cmsRight}{bottom}}{\providecommand{\cmsRight}{right}}
\cmsNoteHeader{SUS-11-020} 
\title{Search for new physics in events with same-sign dileptons and b-tagged jets in pp collisions at $\sqrt{s}=7$ TeV}

\date{\today}

\abstract{
A search for new physics is performed using
isolated same-sign dileptons with at least two b-quark jets
in the final state.
Results are based on a 4.98\fbinv sample of proton-proton
collisions at a centre-of-mass energy of 7\TeV collected by the
CMS detector.
No excess above the standard model background is observed.
Upper limits at 95\% confidence level are set on the number of events
from non-standard-model sources.
These limits are used to set constraints on a number of new physics models.
Information on acceptance and efficiencies are also provided so that the
results can be used to confront additional models in an approximate way.
}

\hypersetup{%
pdfauthor={CMS Collaboration},%
pdftitle={Search for new physics in events with same-sign dileptons and b-tagged jets in pp collisions at sqrt(s) = 7 TeV},%
pdfsubject={CMS},%
pdfkeywords={CMS, physics, supersymmetry}}

\maketitle 

\newcommand{\met}{\ETslash}
\newcommand{\metgen}{\ensuremath{E_{\mathrm{T}}^\text{gen}\hspace{-1.8em}/\kern0.45em}~}
\newcommand{\mll}{\ensuremath{m_{\ell\ell}}}
\newcommand{\intLumi}{\ensuremath{4.98\fbinv}}
\providecommand{\PQu}{\cPqu\xspace}
\providecommand{\PAQu}{\cPaqu\xspace}
\providecommand{\PQb}{\cPqb\xspace}
\providecommand{\PAQb}{\cPaqb\xspace}
\providecommand{\PAQt}{\cPaqt\xspace}
\providecommand{\PQt}{\cPqt\xspace}
\providecommand{\PAp}{\Pap\xspace}
\providecommand{\PZpr}{\cPZpr\xspace}
\providecommand{\PGm}{\Pgm\xspace}
\providecommand{\PGmp}{\Pgmp\xspace}
\providecommand{\PBd}{\ensuremath{\cmsSymbolFace{B_{d}}}\xspace}
\providecommand{\PABd}{\ensuremath{\cmsSymbolFace{\overline{B}_{d}}}\xspace}
\newcommand{\pp}{\Pp{}\Pp}
\newcommand{\ppbar}{\Pp{}\PAp}
\newcommand{\uubar}{\PQu{}\PAQu}
\newcommand{\tbartbar}{\PAQt{}\PAQt}
\section{Introduction\label{sec:introduction}}
We present a search for anomalous production
of events
with two like-sign isolated leptons (\Pe\ or \Pgm) and \PQb-quark jets.
In proton-proton collisions at the Large Hadron Collider (LHC)
such events from standard model (SM) processes are rare; their anomalous
production would be an indication of new physics.
While in general the hadronic jets in new physics processes can originate
from gluons or light flavour quarks, there is a range of well-established models predicting
the presence of two to four \PQb-quark jets in such events.
These appear in signatures of supersymmetry (SUSY)
where bottom- and top-quark superpartners are lighter than
other squarks~\cite{naturalness1,naturalness2,naturalness3,naturalness4,Csaki:2012fh},
enhancing
the fraction of strongly produced
SUSY particles resulting in top and bottom quarks in the final states.
Here, the signatures with two like-sign leptons, \PQb-quark jets and
missing transverse energy
correspond to strongly produced SUSY processes with multiple
W bosons appearing in the decay chains, either from top quarks or charginos.
In addition to SUSY processes,
the existence of a \zp-boson with flavour-violating
\cPqu--\cPqt~quark coupling~\cite{fcnczprime,Buckley}
would lead to like-sign top pair production,
\PQu\!\PQu\ $\to$ \PQt\!\PQt\ via $\zp$ exchange,
at the LHC.
Such a boson has been proposed to explain the top-quark pair
forward-backward production asymmetry observed at
the Tevatron~\cite{d0:fwtop,cdf:fwtop1,cdf:fwtop2}.
A similar topology is expected in models of
maximal flavour violation (MxFV)~\cite{mxflv1,mxflv2,mxflv3}.

Experimentally, events with two isolated like-sign leptons and jets,
selected without \PQb-quark jet identification (\PQb-tagging),
are dominated by \ttbar\ production~\cite{sspaper2010,sspaper2011},
with one lepton from \PW\-decay and the other
lepton
from the semileptonic decay of a \PQb quark.
In a same-sign dilepton selection the requirement of
at least two b-tagged jets
strongly suppresses the \ttbar\ background,
since the two \PQb quarks
in \ttbar\ are very unlikely to produce three distinct objects, \ie,
two b-tagged jets and one isolated high transverse momentum (\pt) lepton.

The search is
performed on a data set corresponding to an integrated luminosity of \intLumi\
collected by the Compact Muon Solenoid (CMS)~\cite{JINST} detector in proton-proton collisions
at $\sqrt{s}=7\TeV$ delivered by the LHC in 2011.
This work relies heavily on the event selections and background
estimation methods of the previous CMS inclusive same-sign dilepton searches
not requiring b-tagged jets in the final
state~\cite{sspaper2010,sstop,sspaper2011}.
Compared with the most recent analysis~\cite{sspaper2011}, a more stringent
isolation requirement is applied to further suppress backgrounds with misidentified
leptons. In addition, the lepton transverse momenta are required to
be above 20\GeV, as is
typical for leptons from \PW\ decays that are expected to be
present
in the signals of interest.
The rest of the data analysis is unchanged.

The search described in this paper is based on the comparison of the number
of observed events with expectations from SM processes.
A loose baseline selection is defined first. Selections with tighter requirements
on the missing transverse energy (\MET) and on the scalar sum of jet \pt (\HT) are then used
to provide better sensitivity to potential signal models.

Since we find no excess of events over the SM background prediction,
we provide a recipe to set limits on any model
with same-sign dileptons, missing transverse energy,
and \PQb-quark jets.
The recipe relies on efficiency functions
to be used to emulate the selection efficiencies for leptons, jets,
and \MET.
These functions can then be applied to a signal simulated
at the matrix-element level.

As a reference, we also provide constraints on several models
representative of this topology.
The signal topologies with two \PQb-quark jets in the final states are:
like-sign top quark production
in the \zp\ model~\cite{fcnczprime} and in the MxFV model~\cite{mxflv3};
production of two bottom squarks each decaying as $\sBot_{1}\to\cPqt\chim_1$.  In the
latter case
$\chim_1\to\PWm\chiz_1$, where $\chiz_1$ is the lightest supersymmetric particle (LSP).
The topologies with more than two \PQb-quark jets are:
$\sGlu\sGlu$ or $\sGlu\sBot$, with $\sGlu\to\sBot_1\cPaqb$
and $\sBot_1\to\cPqt\chim_1$, as above;
$\sGlu\sGlu$ with both gluinos giving a $\cPqt\cPaqt\chiz_1$
final state with an intermediate virtual or on-shell top squark.

\section{CMS detector\label{sec:detector}}
The central feature of the CMS apparatus is a superconducting solenoid,
of 6\unit{m} internal diameter, providing a field of 3.8\unit{T}.
CMS uses a right-handed coordinate system, with the origin defined to be the nominal interaction point,
the $x$ axis pointing to the center of the LHC ring,
the $y$ axis pointing up (perpendicular to the LHC plane),
and the $z$ axis pointing in the anticlockwise beam direction.
The polar angle $\theta$ is measured from the positive $z$ axis and the azimuthal angle $\phi$
is measured in the $x$-$y$ (transverse) plane.
The pseudorapidity $\eta$ is defined as $\eta = - \ln{(\tan{\theta/2})}$.
Within the field volume are the silicon pixel and strip tracker, the crystal electromagnetic calorimeter (ECAL)
and the brass/scintillator hadron calorimeter.
Muons are measured in gas-ionization detectors embedded in the steel return yoke.
Full coverage is provided by the tracker, calorimeters, and the muon detectors within $\abs{\eta}< 2.4$.
In addition to the barrel and endcap detectors up to $\abs{\eta}=3$,
CMS has extensive forward calorimetry reaching $\abs{\eta} \lesssim 5$.
A more detailed description can be found in Ref.~\cite{JINST}.

\section{Event selection\label{sec:selections}}
Dilepton events used in the analysis are selected by the CMS trigger system
if there are at least two leptons (electrons
or muons) reconstructed online.
The trigger selects pairs of leptons above adjustable
thresholds on \pt for muons and $\ET$ for electrons, where
$\ET$ is defined as the energy measured in the ECAL projected
on the transverse plane.
For dielectrons and electron-muon events the thresholds are
17\GeV on the first lepton and 8\GeV on the second lepton.
For dimuon events the requirements on \pt\ for
the higher (lower) threshold changed as the luminosity increased during
data taking from 7~(7)\GeV, to 13~(8)\GeV, and finally reaching 17~(8)\GeV.

Electron candidates are reconstructed
using
measurements provided by the tracker
and the ECAL~\cite{EGMPAS}.
Muon candidates are reconstructed
using a combination of measurements in the silicon tracker and the muon
detectors~\cite{MUOPAS}.
Two leptons of the same sign, $\pt>20\GeV$, and $\abs{\eta}<2.4$,
are required in each event.
Electron candidates
in the transition region between
the barrel and endcap calorimeters ($1.442 < \abs{\eta} < 1.566$) are
not considered in the analysis.
The two leptons
must be consistent with originating from the same collision vertex.
Additional identification requirements are
applied to suppress backgrounds in the same way as in the inclusive
same-sign dilepton analysis~\cite{sspaper2011}.
The isolation requirement is applied on a scalar sum of the track \pt
and calorimeter \ET\ measurements,
computed in a cone of
$\Delta R \equiv \sqrt{(\Delta\eta)^2 + (\Delta\phi)^2}< 0.3$ relative to the lepton
candidate momentum.
This sum must be less than $0.1 \pt$ of the candidate itself.
The two lepton candidates are required to have an invariant mass $m(\ell\ell)$ above 8\GeV to suppress backgrounds
from \PQb-hadron decays.
Events with any third lepton with $\pt>10\GeV$ and isolation sum below $0.2\pt$ are rejected if
this lepton forms an opposite-sign same-flavour pair having
$76\GeV < m(\ell\ell) < 106\GeV$
with either of the selected leptons.
This requirement suppresses the diboson \PW\cPZ\ background.

Jets and missing transverse energy are
reconstructed by the particle-flow
algorithm~\cite{PFPAS,JES,MET}.
Jets are clustered using the anti-$k_\mathrm{T}$ algorithm~\cite{Cacciari:2008gp}
with a distance parameter $R=0.5$.
Jet energies are corrected by subtracting the average contribution from particles
from other proton-proton collisions in the same beam crossing (pileup)
and by
correcting the jet momentum to better reflect the true total momentum
of the particles in the jet~\cite{JES}.
At least two jets with $\pt>40\GeV$ and $\abs{\eta}< 2.5$ are required in each event.
The baseline selection places no requirement on the magnitude of the \MET vector,
computed as the negative of the vector
sum of all particle-flow candidate momenta in the transverse plane.

At least two of the selected jets with $\abs{\eta}< 2.4$ are required to be
\PQb-tagged using the simple secondary vertex tagger at a medium operating point (SSVHEM)~\cite{CMS-PAS-BTV-11-002,CMS-PAS-BTV-11-003}.
This \PQb-tagging algorithm requires the reconstruction of a secondary vertex, with at least two associated tracks,
displaced from the primary collision vertex.  The algorithm has an efficiency between 40--65\% for \PQb-quark jets with $\pt>40\GeV$ and a misidentification rate for
light-quark jets of a few percent, increasing with the transverse momentum.

Events passing the selections described above constitute the baseline
same-sign dilepton sample.
There are 10 such events observed in data: 3 $\Pe\Pe$, 2 $\Pgm\Pgm$, and 5 $\Pe\Pgm$.

\section{Background estimation\label{sec:backgrounds}}
There are three distinct background contributions to this search: events with one or two
``fake'' leptons, rare SM processes that yield events with two isolated same-sign leptons,
and events with opposite-sign lepton pairs with a lepton charge misreconstructed (``charge-flips'').
Here we define the term ``fake lepton'' to refer to a lepton from heavy
flavour decay, an electron from
unidentified photon conversion, a muon from meson decays in flight, or a hadron
misidentified as a lepton.  The backgrounds,
which are further discussed below, are
estimated using the same techniques as in the
inclusive analysis~\cite{sspaper2011,sspaper2010}: the fake and
charge-flip backgrounds are
estimated from control data samples, while the rare SM
backgrounds are determined from simulation.

The background from fakes is estimated from events where one or both leptons
fail the tight isolation and identification selection,
but still pass a looser selection.
Counts of events in this control sample are weighted by the expected ratio
(``tight-to-loose'', or TL ratio) of
the rate of fake leptons passing the selection to that of those failing it.
This TL ratio
is measured
as a function of lepton type, \pt, and $\eta$,
in a data sample of events
with a single lepton candidate and a well separated jet (``away-jet'').
After vetoing \cPZ\ candidates and suppressing leptons from
\PW\ decays by requiring small \MET and transverse mass, the leptons
in this sample are predominantly fakes.
The systematic effects on the method to estimate events with fake leptons
arise from differences in kinematics and sample composition between the sample
where the TL ratio is measured and the sample where it is applied.
The systematic uncertainty on the method is taken to be 50\%.
This uncertainty is based on tests of the ability of this method
to predict the same-sign dilepton background in simulated \ttbar\ events;
it is also based
on the observed variations of the TL ratio as a function of the
$\pt$ threshold of the away jet and the addition of a \PQb-tag requirement
on that jet.

The baseline sample is estimated to have $1.5\pm 1.1$, $0.8\pm0.5$, and $2.4\pm1.4$ events
with fake leptons
in the $\Pe\Pe$,
$\Pgm\Pgm$, and $\Pe\Pgm$ final states, respectively.  These uncertainties
include a statistical uncertainties based on the number of events passing
the loose lepton selection,
as well as the 50\% systematic uncertainty.

As mentioned above, we estimate, from simulation, the contribution to the event count from
rare SM processes yielding isolated high-\pt same-sign dileptons and jets.
Events are generated with the
\MADGRAPH~\cite{MADGRAPH4} event
generator
and then passed on to \PYTHIA~\cite{PYTHIA}
for parton shower and hadronization.
The generated events
are processed by the CMS event simulation and the same chain
of reconstruction programs as is
used for collision data.  As determined from simulation, we find that
background events from \ttbar\!\PW\ and \ttbar\!\cPZ\ production represent more than 90\%
of all the genuine same-sign dilepton backgrounds.
Other processes considered include production of diboson (\PW\cPZ, \cPZ\cPZ, same-sign \PW\PW)
and triboson (combinations of \PW\ and \cPZ) final states.
Compared to the inclusive analysis~\cite{sspaper2011}, these backgrounds are strongly suppressed
by the \PQb-tagging requirement.
Backgrounds like (\PW/\cPZ)\cPgg\ and \ttbar\!\cPgg\ are considered as well to
simulate events with a photon converting in the tracker material and misidentified as an electron.
Their contribution is negligibly small.
A
conservative
systematic uncertainty of 50\% is assigned to the total
number of background events from simulation, since these are
rare SM processes which have yet to be observed.
The production cross sections used to normalize the dominant
\ttbar\!\PW\ and \ttbar\!\cPZ\ contributions
are 0.16\unit{pb}~\cite{Campbell:2012dh}
and 0.14\unit{pb}~\cite{ttzNLO,Garzelli:2011is}, respectively.
In the baseline sample the simulated rare
SM backgrounds are determined to
contribute $0.9\pm0.5$, $1.1\pm0.6$, and $2.0\pm1.0$
events in the $\Pe\Pe$, $\Pgm\Pgm$, and $\Pe\Pgm$ final states, respectively.

Events with opposite-sign lepton pairs where one of the leptons
has an incorrectly measured charge (charge-flip) contribute to the same-sign dilepton
sample.  The charge-flip probability for muons is of order
$10^{-4}$--$10^{-5}$
and can be neglected.  In contrast,
this probability
for electrons
from
\PW\ or
\cPZ\ decay is estimated in simulation to be about $10^{-3}$.
The number of same-sign events due to charge-flips is given by the
number of opposite-sign events
passing the same selections with a weight applied to
each electron corresponding
to its charge misidentification probability.
We determine this probability in simulation as a function
of electron \pt and $\eta$.
The method is tested in data by using the \cPZ\ $\to \EE$ sample and the
probability mentioned above to predict
the number of $\Pe^{\pm}\Pe^{\pm}$ events with invariant mass consistent with the
\cPZ\ mass.  This prediction is found to be in good agreement with the number of
events of this type in data.
A systematic uncertainty of 20\% is estimated for this method based on variation in
the average charge misidentification rate between typical lepton momenta  in  \cPZ\ and  \ttbar\ events.
In the baseline sample the charge-flip contribution is estimated to be $0.8\pm0.2$ and $0.6\pm0.1$ events in the $\Pe\Pe$
and $\Pe\Pgm$ final states, respectively.

\section{Search results}
\label{sec:yields}
After the basic selection described in Section~\ref{sec:selections},
we define several ``signal regions'' (SR)  with increasing requirements
on $\HT$ and \MET with respect to the baseline selection.
These requirements improve the sensitivity
to new physics models with high mass scales
and/or high
\MET from, \eg, high $\pt$ non interacting particles, such
as LSPs in SUSY models.  We also define a SR
with minimal requirements on $\HT$ and $\MET$ but allowing only
for positive leptons.  This region is designed to be sensitive to
$\Pp \Pp \to \PQt \PQt$ production (in most models $\Pp \Pp \to \PAQt \PAQt$
is suppressed with respect to $\Pp \Pp \to \PQt \PQt$ since
at the parton level these processes originate from $\PAQu \PAQu$
and $\PQu \PQu$ initial states, respectively).
Additionally, we define a SR with moderate $\HT$ and $\MET$
requirements and three or more \PQb-tagged jets.
This region can improve the sensitivity to models of
new physics with several
($\geq 3$) \PQb quarks in the final state.  However, for the models
considered here (Section~\ref{sec:model}) we find that inclusion
of this region does
not improve the sensitivity.  This is because the increase in
efficiency due to the looser $\HT$ and $\MET$ requirements
does not compensate for the efficiency loss associated with
the requirement of a third \PQb-tag.  Finally, we define a SR with a high $\HT$
requirement and no $\MET$ requirement.
This region is designed to enhance sensitivity to models
with R-parity violating SUSY~\cite{rpv}
with~\cite{rpvnomet} or without~\cite{rpvwmet,Dreiner:2012mn}
leptonically decaying \PW\ bosons (the latter type of events
have no intrinsic \MET from undetected particles).

\begin{table}
\topcaption{\label{tab:SR} A summary of the results of this search.
For each signal region (SR),
we show its most distinguishing kinematic requirements, the prediction
for the three background  (BG) components as well as the total,
the event yield, and the observed 95\% confidence
level upper limit on the number of non-SM events ($N_{UL}$) calculated
under three different
assumptions for the event efficiency uncertainty (see text for details).
Note that the count of the number of jets on the first line of
the table includes both tagged and untagged jets.}
\tabcolsep 2.7pt
\begin{scriptsize}
\begin{tabular}{|l|c|c|c|c|c|c|c|c|c|}
\hline
 & SR0 & SR1 &  SR2 & SR3 & SR4 & SR5 & SR6 & SR7 & SR8 \\
\hline
No. of jets & $\geq 2$ & $\geq 2$ &  $\geq 2$ &  $\geq 2$ &  $\geq 2$ &  $\geq 2$ &  $\geq 2$ &  $\geq 3$ & $\geq 2$\\
No. of \PQb-tags & $\geq 2$ & $\geq 2$ &  $\geq 2$ &  $\geq 2$ &  $\geq 2$ &  $\geq 2$ &  $\geq 2$ &  $\geq 3$ & $\geq 2$\\
Lepton charges & $++/--$ & $++/--$ & $++$ & $++/--$ & $++/--$ & $++/--$ & $++/--$ & $++/--$ & $++/--$ \\
$\MET$ & $> 0$ \GeV & $> 30$ \GeV & $> 30$ \GeV & $> 120$ \GeV & $> 50$ \GeV & $> 50$ \GeV & $> 120$ \GeV & $> 50$ \GeV & $> 0$ \GeV\\
$\HT$ & $> 80$ \GeV & $> 80$ \GeV & $> 80$ \GeV & $> 200$ \GeV & $> 200$ \GeV & $> 320$ \GeV & $> 320$ \GeV & $> 200$ \GeV & $> 320$ \GeV \\
\hline
Charge-flip BG & $1.4 \pm 0.3$ & $1.1 \pm 0.2$ & $0.5 \pm 0.1$ & $0.05 \pm 0.01$ & $0.3 \pm 0.1$ & $0.12 \pm 0.03$ & $0.03 \pm 0.01$ & $0.008 \pm 0.004$ & $0.20 \pm 0.05$ \\
Fake BG & $4.7 \pm 2.6$ & $3.4 \pm 2.0$ &  $1.8 \pm 1.2$ &  $0.3 \pm 0.5$ &  $1.5 \pm 1.1$ &  $0.8 \pm 0.8$ &  $0.15 \pm 0.45$ &  $0.15 \pm 0.45$ & $1.6 \pm 1.1$ \\
Rare SM BG & $4.0 \pm 2.0$ & $3.4 \pm 1.7$ & $2.2 \pm 1.1$ & $0.6 \pm 0.3$ & $2.1 \pm 1.0$ & $1.1 \pm 0.5$ & $0.4 \pm 0.2$ & $0.12 \pm 0.06$ & $1.5 \pm 0.8$\\
\hline
Total BG  & $10.2 \pm 3.3$ & $7.9 \pm 2.6$ & $4.5 \pm 1.7$ & $1.0 \pm 0.6$ & $3.9 \pm 1.5$ & $2.0 \pm 1.0$ & $0.6 \pm 0.5$ & $0.3 \pm 0.5$ & $3.3 \pm 1.4$ \\
Event yield & 10 & 7 & 5 & 2 & 5 & 2 & 0 & 0 & 3 \\
\hline
$N_{UL}$ (12\% unc.) & 9.1 & 7.2 & 6.8 & 5.1 & 7.2 & 4.7 & 2.8 & 2.8 & 5.2  \\
$N_{UL}$ (20\% unc.) & 9.5 & 7.6 & 7.2 & 5.3 & 7.5 & 4.8 & 2.8 & 2.8 & 5.4 \\
$N_{UL}$ (30\% unc.) & 10.1 & 7.9 & 7.5 & 5.7 & 8.0 & 5.1 & 2.8 & 2.8 & 5.7 \\
\hline
\end{tabular}
\end{scriptsize}
\end{table}

\begin{figure}[hbtp]
\begin{center}
\includegraphics[width=\linewidth]{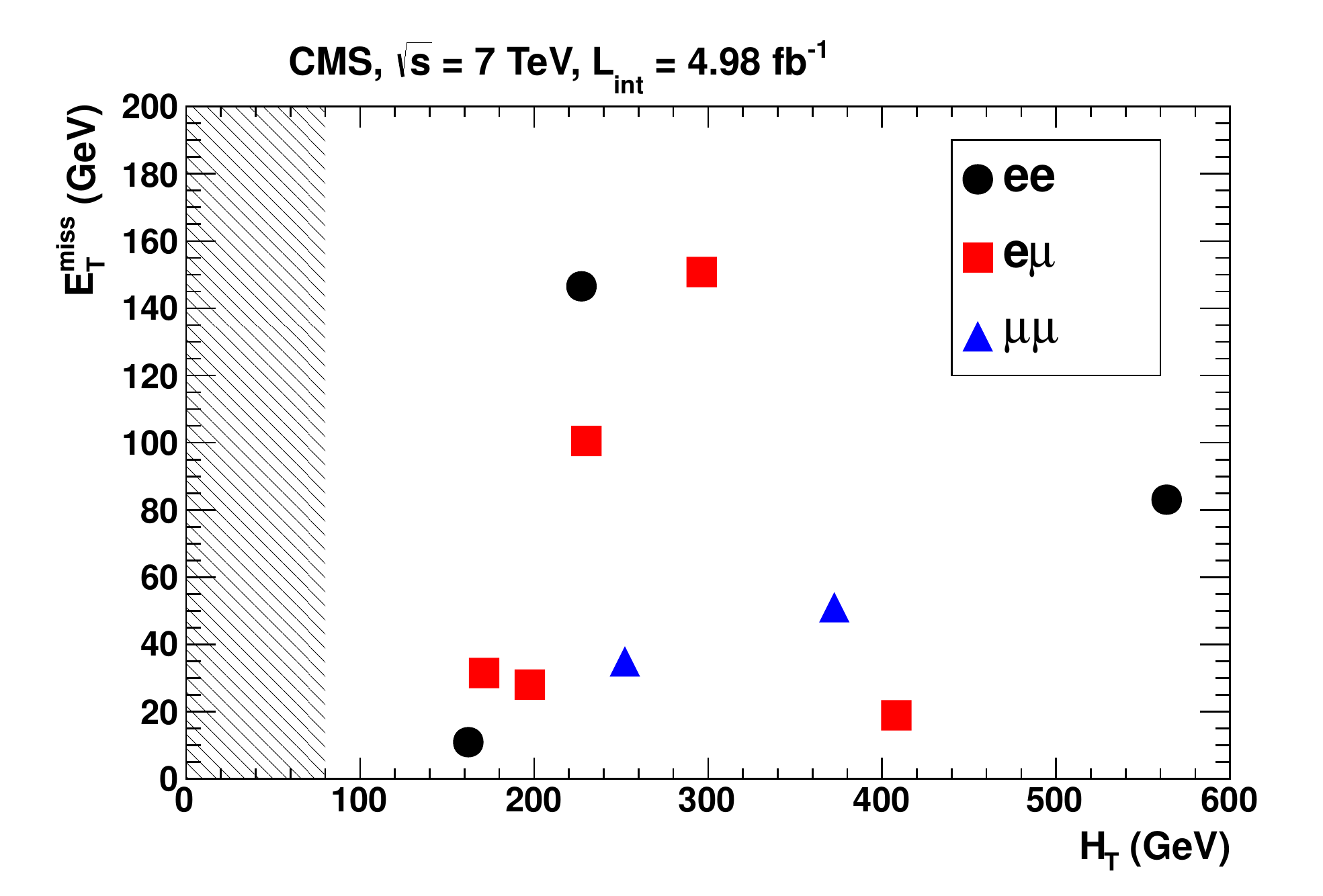}
\includegraphics[width=0.49\linewidth]{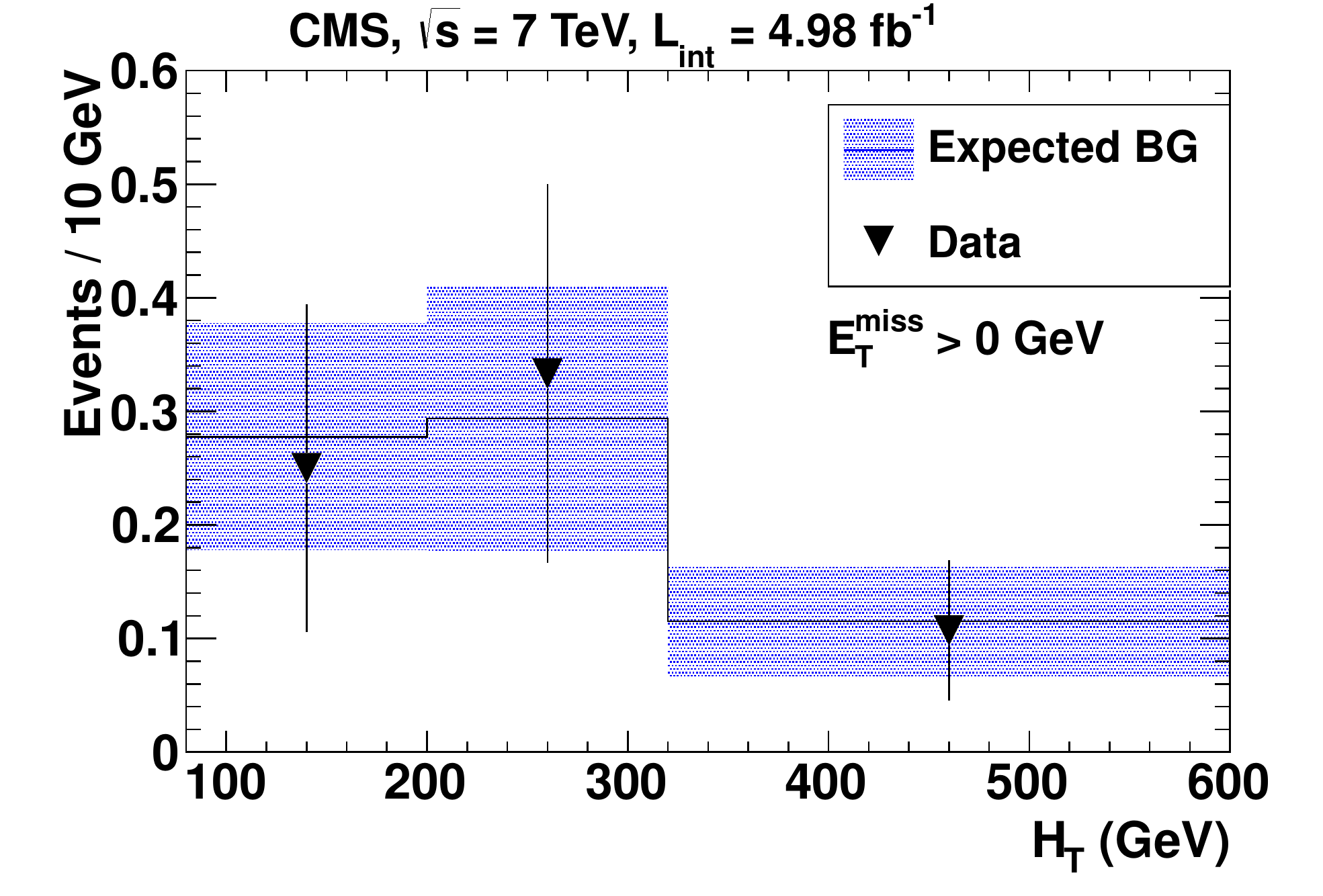}
\includegraphics[width=0.49\linewidth]{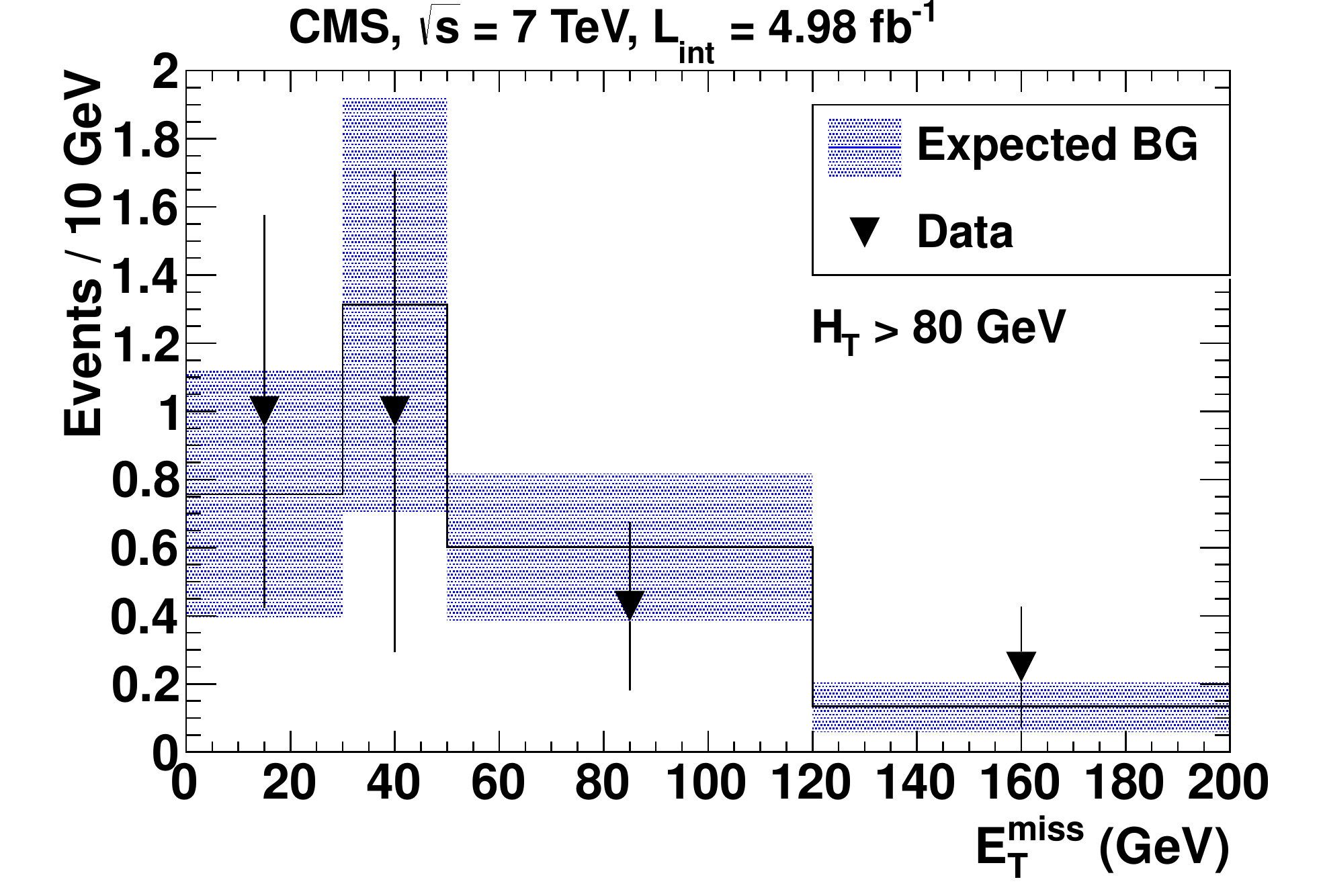}
\topcaption{\label{fig:htmet}
Top plot: distribution of \MET vs. $\HT$ for the 10 events in the
baseline region (SR0).
Note that the $\geq 2$ jets requirement in SR0 implies $\HT > 80$\GeV.
Bottom left plot: projection of the scatter plot on the $\HT$ axis.
Bottom right plot: projection of the scatter plot on the \MET axis.
For the one-dimensional distributions, the number of events in each bin is scaled appropriately to reflect units of events per
10\GeV and is compared with the background (BG)
predictions, with their uncertainties.}
\end{center}
\end{figure}

The definitions of the signal regions, the data event yields, and the
expected backgrounds calculated for each SR,
are summarized in Table~\ref{tab:SR}.
Distributions of $\HT$ and \MET are also displayed
in Fig.~\ref{fig:htmet} for the baseline selection.
Note that SR0
corresponds to the baseline event selection of Section~\ref{sec:selections}.
The event yields are consistent with the background predictions.
In Table~\ref{tab:SR} we also show the 95\% confidence level
observed upper limit ($N_{UL}$)
on the number
of non-SM events calculated using the CL$_s$ method~\cite{Junk:1999kv,ATL-PHYS-PUB-2011-011}
under three different assumptions for the
signal efficiency uncertainty.
This uncertainty is discussed in Section~\ref{sec:eff}.

\section{Efficiencies and associated uncertainties}
\label{sec:eff}

\begin{figure}[h]
\begin{center}
\includegraphics[width=0.49\linewidth]{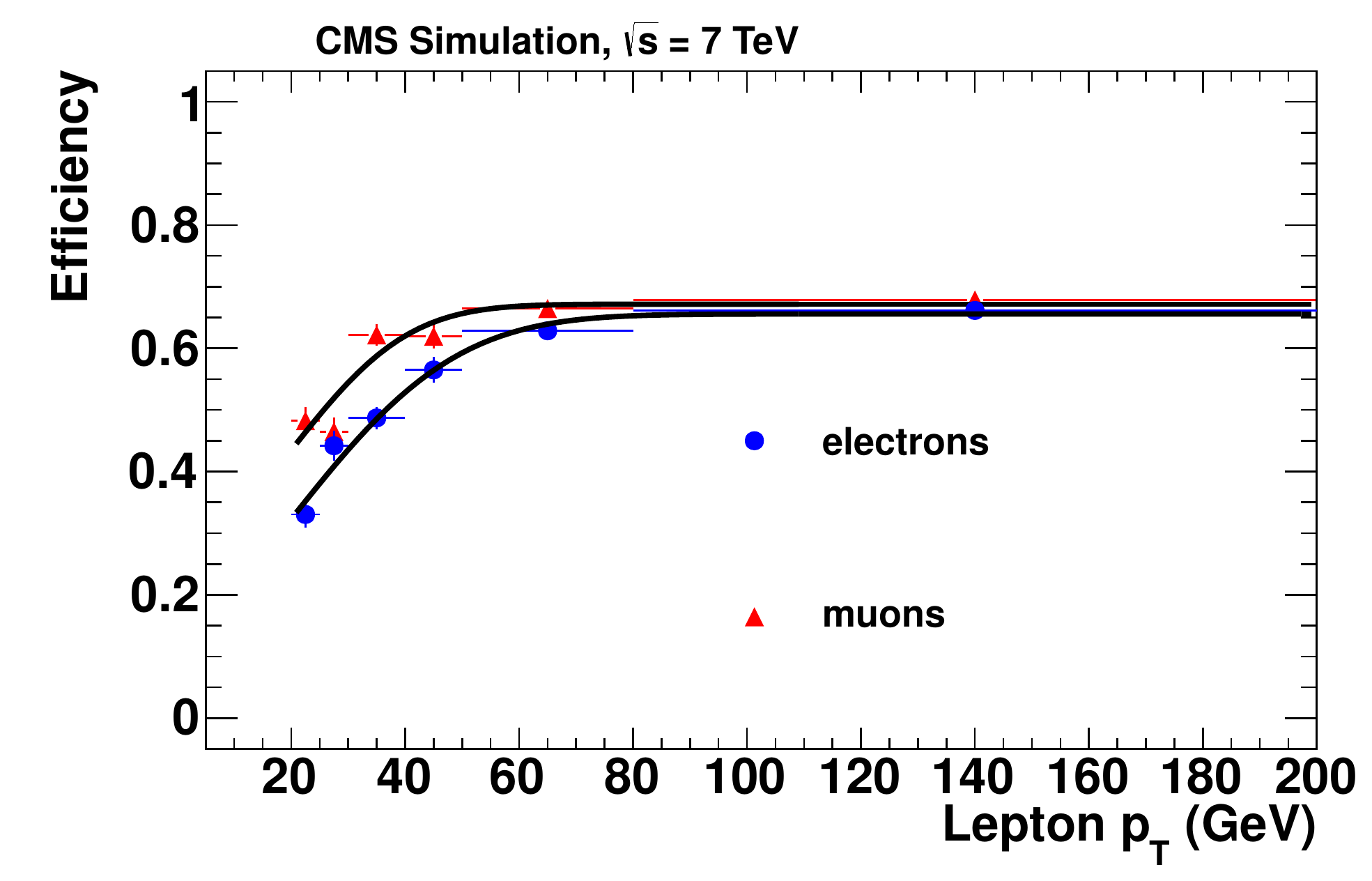}
\includegraphics[width=0.49\linewidth]{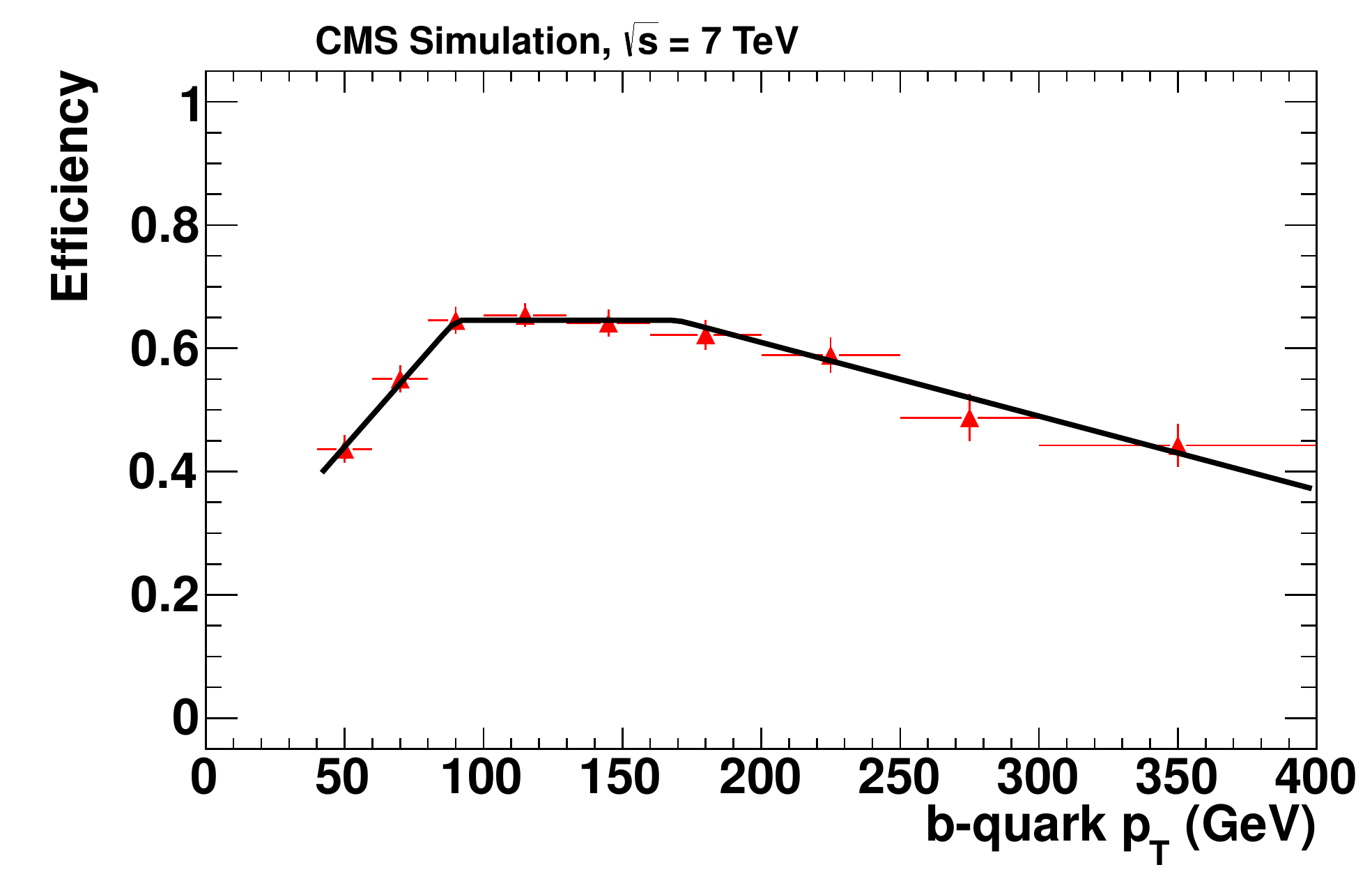}
\caption{\label{fig:efficiencies}
Lepton selection efficiency as a function of $\pt$ (left);
b-jet tagging efficiency as a function of the \PQb quark $\pt$ (right).
}
\end{center}
\end{figure}

Events in this analysis are collected with dilepton triggers.
The efficiency of the trigger is measured to be $99 \pm 1\%$
($96 \pm 3\%$) per electron (muon) in the range $\abs{\eta} < 2.4$.
The efficiency of the lepton identification and isolation
requirements, as determined using a sample of simulated events
from a typical SUSY scenario (the LM6 point of
Ref.~\cite{PTDR2}), is displayed in
Fig.~\ref{fig:efficiencies}.  Studies of large data
samples of $\cPZ \to \Pe\Pe$ and $\cPZ \to \Pgm\Pgm$ events indicate that
the simulation reproduces the efficiencies of the identification
requirements to better than 2\%~\cite{EGMPAS,MUOPAS}.
The efficiency
of the isolation requirement on leptons in
\cPZ\ events is also well reproduced by the simulation.  However,
this efficiency depends on the hadronic activity in the event,
and is typically 10\% lower in SUSY events with hadronic cascades
than in \cPZ\ events.
To account for this variation, we take a
5\% systematic uncertainty per lepton in the acceptance of signal events.

The \PQb-tagging efficiency on simulated data is also shown in
Fig.~\ref{fig:efficiencies} for \PQb quarks of $\abs{\eta} < 2.4$
and $\pt > 40$\GeV.
Study of a variety of control samples indicate that for collision
data this
efficiency needs to be reduced by a factor of 0.96, independent
of $\pt$.  This factor is applied to the simulation of
possible new physics signals, \eg, all the models
of Section~\ref{sec:model}.  The systematic uncertainty on the
\PQb-tagging efficiency is 4\% (15\%) for jets of $\pt < 240$\GeV
($\pt > 240$\GeV).

The energies of jets in this analysis are known to
7.5\% (not all the corrections described in
Ref.~\cite{JES} were applied, since they have little
impact on the sensitivity of this search).
The uncertainty on the jet energy scale
has an effect on the efficiencies of the jet multiplicity, $\HT$, and
$\MET$ requirements.  The importance of these effects depends on the
signal region and the model of new physics.
For example, for the $\zp$ model of Section~\ref{sec:ttmodel}, the
uncertainty on the acceptance of the SR2 requirements due to the imperfect
knowledge of the jet energy scale is 8\%.
In general, models
with high hadronic activity and high \MET are less affected
by this uncertainty.

The total uncertainty on the acceptance is in the 12--30\% range.
Finally, there is a 2.2\% uncertainty on the yield of events from
any new physics model due to the uncertainty in the luminosity
normalization~\cite{CMS-PAS-SMP-12-008}.

\section{Information for model testing}
\label{sec:outreach}

We have described a signature based search that finds no evidence
for physics beyond the SM.
In Section~\ref{sec:model}
we will use our results to put bounds on the parameters of a number of
models of new physics.
Here we present additional information
that can be used to confront other models of new physics in an approximate
way by generator-level studies that compare the expected number of
events
with the upper limits from
Table~\ref{tab:SR}.

The values of $N_{UL}$
for the different signal regions are given
in Table~\ref{tab:SR}
under different assumptions for the efficiency uncertainty.  This is because,
as discussed in Section~\ref{sec:eff}, this uncertainty depends on the
model under test.  The dependence of $N_{UL}$ on the acceptance
uncertainty is not very strong. Thus, for the purpose of generator-level
model testing, the lack of precise knowledge of the uncertainty
does not constitute a significant limitation.

\begin{figure}[h]
\begin{center}
\includegraphics[width=0.49\linewidth]{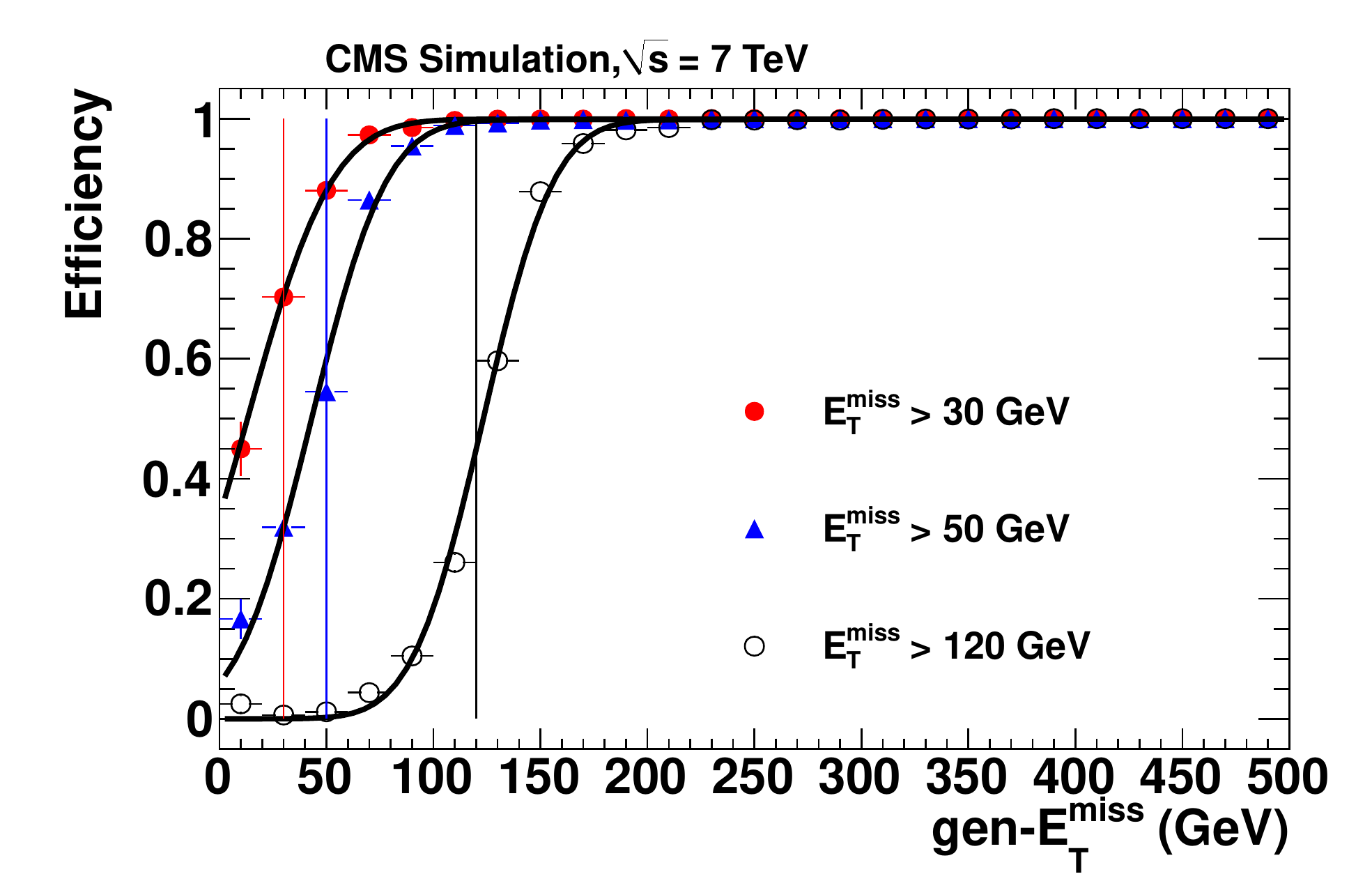}
\includegraphics[width=0.49\linewidth]{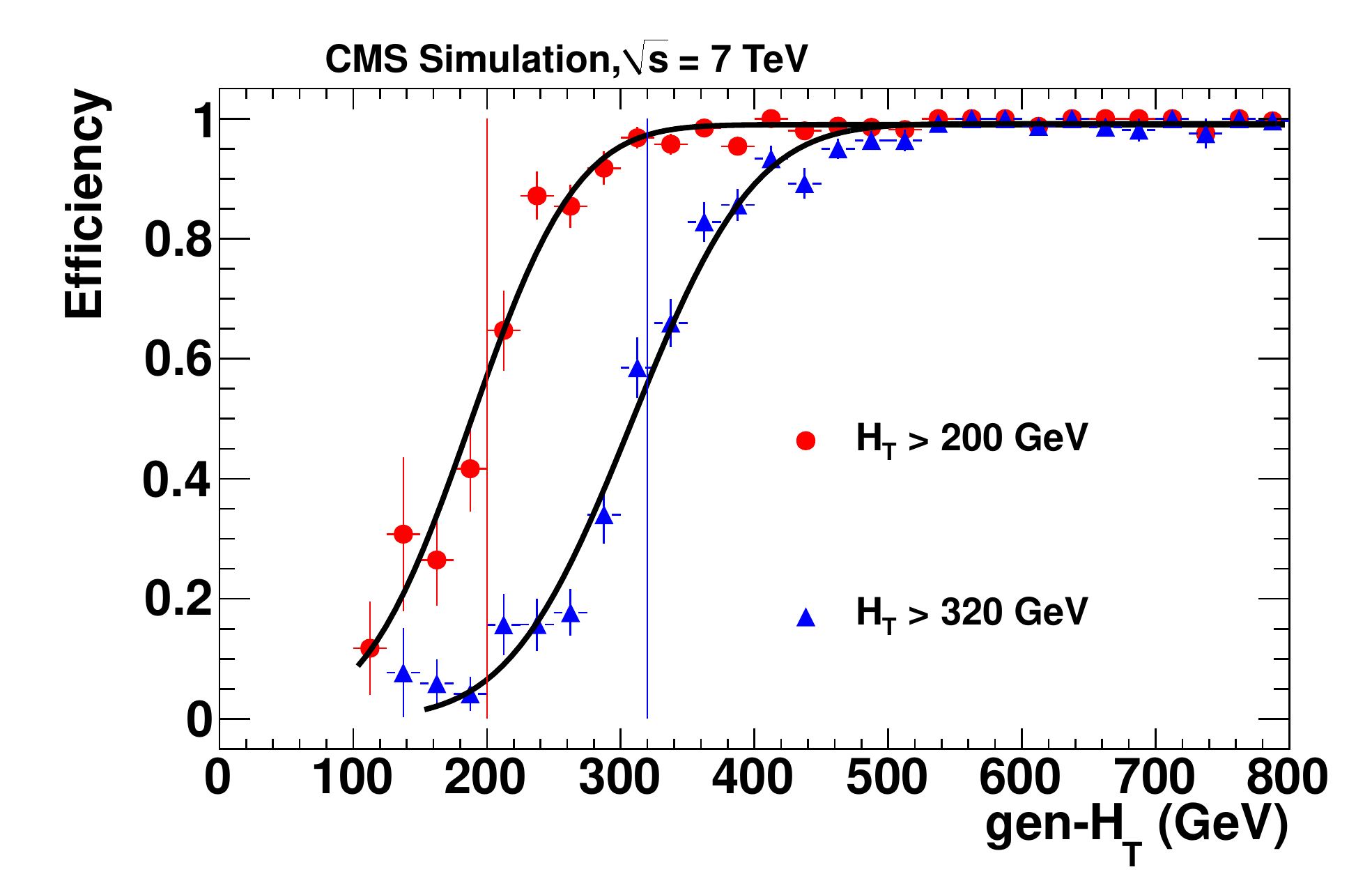}
\caption{\label{fig:efficiencies2}
Efficiency for an event to pass a given reconstructed \MET\ or $\HT$
threshold
as a function of gen-$\MET$ or gen-$\HT$.
The efficiencies are shown for the thresholds
used in defining the signal regions.
}
\end{center}
\end{figure}

The kinematic requirements on jets and leptons
given in Section~\ref{sec:selections}
are the first ingredients of the acceptance calculation for a new model.
Leptons at the hard-scatter level passing the kinematic selection can
be counted, and this count can be corrected for the finite lepton efficiencies shown
in Fig.~\ref{fig:efficiencies}, as well as the trigger efficiencies
given in Section~\ref{sec:eff}.
Similarly, the number of jets in the
event can be approximated by
counting the number of colored final-state partons
of $\pt > 40$\GeV
and $\abs{\eta} < 2.5$ at the hard scatter level.
A generator-level $\HT$ variable, gen-$\HT$, can be calculated by
summing the $\pt$ of all the colored partons from the previous step;
isolated photons and additional leptons of $\pt > 40$\GeV and
$\abs{\eta} < 2.5$  should also be included in the gen-$\HT$
calculation.
Similarly,
a generator-level \MET variable, gen-$\MET$, can be defined
from the vector sum of transverse momenta of all non-interacting particles.
Finally, the number of reconstructed \PQb-quark jets can be obtained
by counting the number of \PQb quarks and
applying the efficiency parametrization of Fig.~\ref{fig:efficiencies},
including the requirements $\pt > 40$\GeV and $\abs{\eta}<2.4$.
The efficiencies of the $\HT$ and \MET requirement after
hadronization and detector simulation
as a function of gen-$\HT$ and gen-$\MET$ for a
typical SUSY scenario are shown
in Fig.~\ref{fig:efficiencies2}.

The lepton efficiency curves of Fig.~\ref{fig:efficiencies}
are parametrized as

\begin{eqnarray}
\epsilon = \epsilon_{\infty} \erf\left ( \frac{\pt - 20\GeV}{\sigma} \right)
        + \epsilon_{20} \left( 1.- \erf\left ( \frac{\pt - 20\GeV}{\sigma} \right) \right),
\label{eq:lepeffFitF}
\end{eqnarray}

\noindent with $\epsilon_{\infty} = 0.66$ (0.67),
$\epsilon_{20} = 0.32$ (0.44), $\sigma = 32$\GeV (23\GeV) for electrons
(muons).

The parametrization of the simulated \PQb-tagging efficiency, also shown in Fig.~\ref{fig:efficiencies},
is $\epsilon = 0.62$ for $90  < \pt < 170$\GeV; at higher
(lower) $\pt$ it decreases linearly with a slope of \linebreak[2]$0.0012\ (0.0051)\,\mathrm{GeV}^{-1}$.

The $\HT$ and \MET turn-on curves as a function of the respective
generator version shown in Fig.~\ref{fig:efficiencies2}
are parametrized as
$0.5 \{\erf[(x - x_{1/2})/\sigma] + 1\}$.
The parameters of the function are summarized in Table~\ref{tab:methtpar}.

\begin{table}[h]
\begin{center}
\topcaption{\label{tab:methtpar} Parameters used in describing the turn-on
curves for $\HT$ and \MET as a function of their generator-level values.
See text for details.}
\begin{tabular}{l|rr|rrr}\hline\hline
Parameter               & \multicolumn{2}{c|}{$\HT$}                      & \multicolumn{3}{c}{\MET} \\
                        &       ${>}200\GeV$     &       ${>}320\GeV$    & ${>}30\GeV$  & ${>}50\GeV$            & ${>}120\GeV$   \\
\hline
$x_{1/2}$               & 188\GeV                & 308\GeV               & 13\GeV       & 43\GeV                 & 123\GeV  \\
$\sigma$                & 88\GeV                 & 102\GeV               & 44\GeV       & 39\GeV                 & 37\GeV \\
\hline\hline
\end{tabular}
\end{center}
\end{table}

For a few of the models of new physics
described in Section~\ref{sec:model},
we have compared the acceptance from the full simulation with the
result of the simple acceptance model described above.
For scenarios with at least two \PQb quarks in the final state,  the two calculations typically agree at the ${\approx}15\%$ level or better.
However, in scenarios where \PQb quarks are rare or where the lepton isolation is significantly different than in a typical SUSY event, the two calculations may vary by ${\approx}30\%$ or more.

\section{Models of new physics}
\label{sec:model}

We use the search results to constrain several specific
models of
new physics.  Signal samples are generated using \PYTHIA\ with
the detector simulation performed using the CMS fast
simulation package~\cite{Abdullin:1328345, CMS-DP-2010-039}.
For each model
considered, we use the simulated signal yields and the background
estimations corresponding to
the signal region that is expected to give the most
stringent limit on the cross section at a given point
in model parameter space.  Cross section limits are computed
using the CL$_s$ method~\cite{Junk:1999kv,ATL-PHYS-PUB-2011-011}
including
systematic uncertainties on lepton efficiency (5\% per lepton),
luminosity (2.2\%),
jet energy scale, and \PQb-tagging efficiency.  These last
two uncertainties
are evaluated at each point in parameter
space, as they depend on the underlying kinematics of the
events.
In addition, the simulated event yields are corrected
for ``signal contamination'', \ie, the oversubtraction
of the fake background that would occur in the presence
of a real signal.  This oversubtraction is caused
by same-sign dilepton events with one lepton passing the
loose selection but failing the final identification or
isolation requirements.
The cross section limits are then used to exclude
regions of model parameter space.

\subsection{Models of \texorpdfstring{$\pp \to \PQt \PQt$}{pp to tt}}
\label{sec:ttmodel}
We consider two models that result in same-sign top-quark pairs
without significant additional hadronic activity or missing transverse
energy.  Limits are set based on the results from SR2.
The kinematic requirements in this region are modest, and are
comparable to those used in the CMS measurements of the $\pp \to \ttbar$
cross section in the opposite-sign dilepton channel~\cite{cmsdil,Khachatryan:2010ez}.
We require only positively charged dileptons, since in the two models
considered $\PQt \PQt$ production dominates over $\PAQt \PAQt$.

The first model is the $\zp$ model of
Ref.~\cite{fcnczprime}, which is proposed as a possible
explanation of the anomalous forward-backward asymmetry
observed at the Tevatron~\cite{cdf:fwtop1,cdf:fwtop2,d0:fwtop}.
This model introduces a new neutral boson
with chiral couplings to u and t quarks.
The relevant term in the Lagrangian is
$\mathcal{L} = \frac{1}{2}g_W f_R \bar{u} \gamma^\mu (1+\gamma^5)t\zp_\mu +
\text{h.c.}$, and the model parameters are $f_R$ and the mass of the $\zp$, $m(\zp)$.
In this model same-sign top pairs are produced predominantly through
$t$-channel $\zp$ exchange in \PQu\!\PQu\ $\to$ \PQt\!\PQt.

The efficiency for $\pp \to \PQt \PQt$
events in the $\zp$ model is calculated
from simulated events, first generated with \MADGRAPH
and then processed by \PYTHIA.
We find an efficiency, including branching fractions,
of $0.23 \pm 0.04$\%,
largely independent of $m(\zp)$.   The
resulting cross section upper limit
is 0.61\unit{pb} at the 95\% confidence level.  This
improves the previous CMS limit~\cite{sstop} by a factor of 27.
This improvement is due to the factor 140 increase in the integrated
luminosity between the two analyses.  The limit scales
faster than the inverse of the square root of the luminosity since
the addition of the \PQb-tag requirement has reduced the background
level by a large factor.
Our limit is a factor of 2.8 more stringent than that reported
by the ATLAS collaboration~\cite{sstopatlas}.

In order to compare with other experiments, we also interpret our
result in terms of an effective four-fermion Lagrangian for 
$\PQu \PQu \to \PQt \PQt$~\cite{cdfth2}:

\begin{eqnarray}
\mathcal{L} & = &
\frac{1}{2}\frac{C_{RR}}{\Lambda^2}
 [\PAQu_R \gamma^\mu \PQt_R][\PAQu_R \gamma_{\mu} \PQt_R] + 
\frac{1}{2}\frac{C_{LL}}{\Lambda^2}
 [\PAQu_L \gamma^\mu \PQt_L][\PAQu_L \gamma_{\mu} \PQt_L]  - \nonumber \\
&  & \frac{1}{2}\frac{C_{LR}}{\Lambda^2}
 [\PAQu_L \gamma^\mu \PQt_L][\PAQu_R \gamma_{\mu} \PQt_R] -
\frac{1}{2}\frac{C'_{LR}}{\Lambda^2}
 [\PAQu_{La} \gamma^\mu \PQt_{Lb}][\PAQu_{Rb} \gamma_{\mu} \PQt_{Ra}] + 
\text{h.c.}
\end{eqnarray}

\noindent where $a$ and $b$ are color indices.  
Note that at large $m(\zp)$ the Lagrangian for the $\zp$ model corresponds to the first
term in the effective  Lagrangian with
$\frac{g_W^2  f_R^2}{m(\zp)^2} = \frac{C_{RR}}{\Lambda^2}$.
In this framework our
limit on $\sigma(\PQt \PQt)$ results in limits
$\frac{C_{RR}}{\Lambda^2}$ or $\frac{C_{LL}}{\Lambda^2} <0.20\TeV^{-2}$
and 
$\frac{C_{LR}}{\Lambda^2}$ or $\frac{C'_{LR}}{\Lambda^2} <0.56\TeV^{-2}$, all
at the 95\% CL.  
These bounds are more stringent than those of
CDF~\cite{cdflimit} and Atlas~\cite{sstopatlas}.

The second model~\cite{mxflv1,mxflv2,mxflv3} has a new scalar
SU(2) doublet $\Phi = (\eta^0,\eta^+)$ that couples the first and third
generation quarks ($q_1,q_3$) via a Lagrangian term
$\mathcal{L} = \xi \Phi q_1 q_3$.  Remarkably,
this
model is largely consistent with constraints from flavour physics.  The
parameters of this ``Maximally Flavour Violating'' (MxFV) model are
the mass of the $\eta^0$ boson and the value of the coupling $\xi$.
In the MxFV model,
same-sign top pairs are produced dominantly in $\PQu \PQu \to \PQt \PQt$ through
$t$-channel $\eta^0$ exchange.  At small values of $\xi$ and
$\eta^0$ mass
$\PQu \Glu  \to \eta^0 \to \PQt \PQt \PQu$ becomes important. The third production
mechanism, $\PQu \PQu \to \eta^0 \eta^0$, is also considered in our analysis.
Signal events in the MXFV model are generated using \MADGRAPH followed
by \PYTHIA for showering and hadronization. The decay widths are computed using the
\textsc{bridge} program~\cite{Meade:2007js}.

\begin{figure}[htb]
\begin{center}
\includegraphics[width=0.49\linewidth]{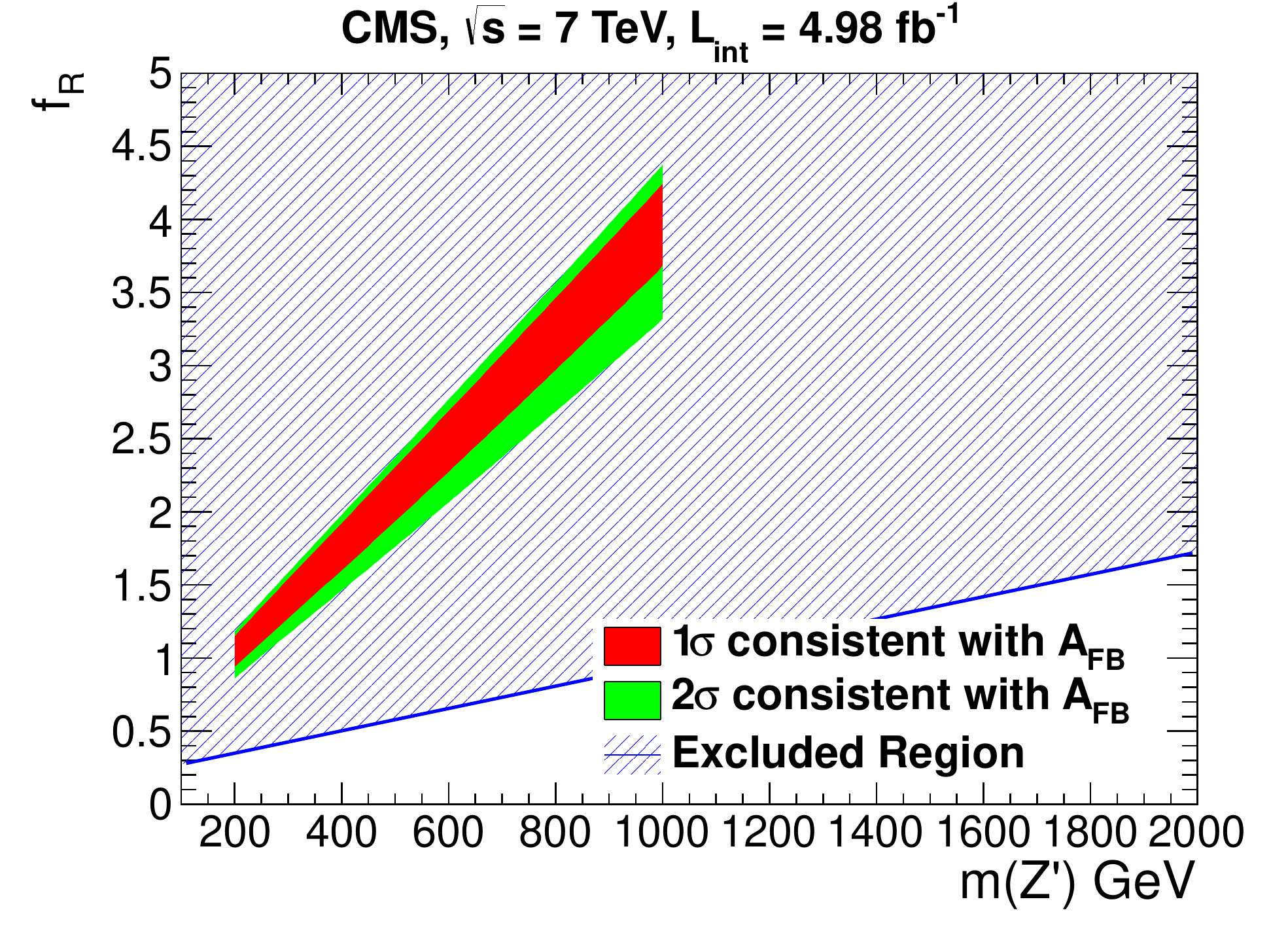}
\includegraphics[width=0.49\linewidth]{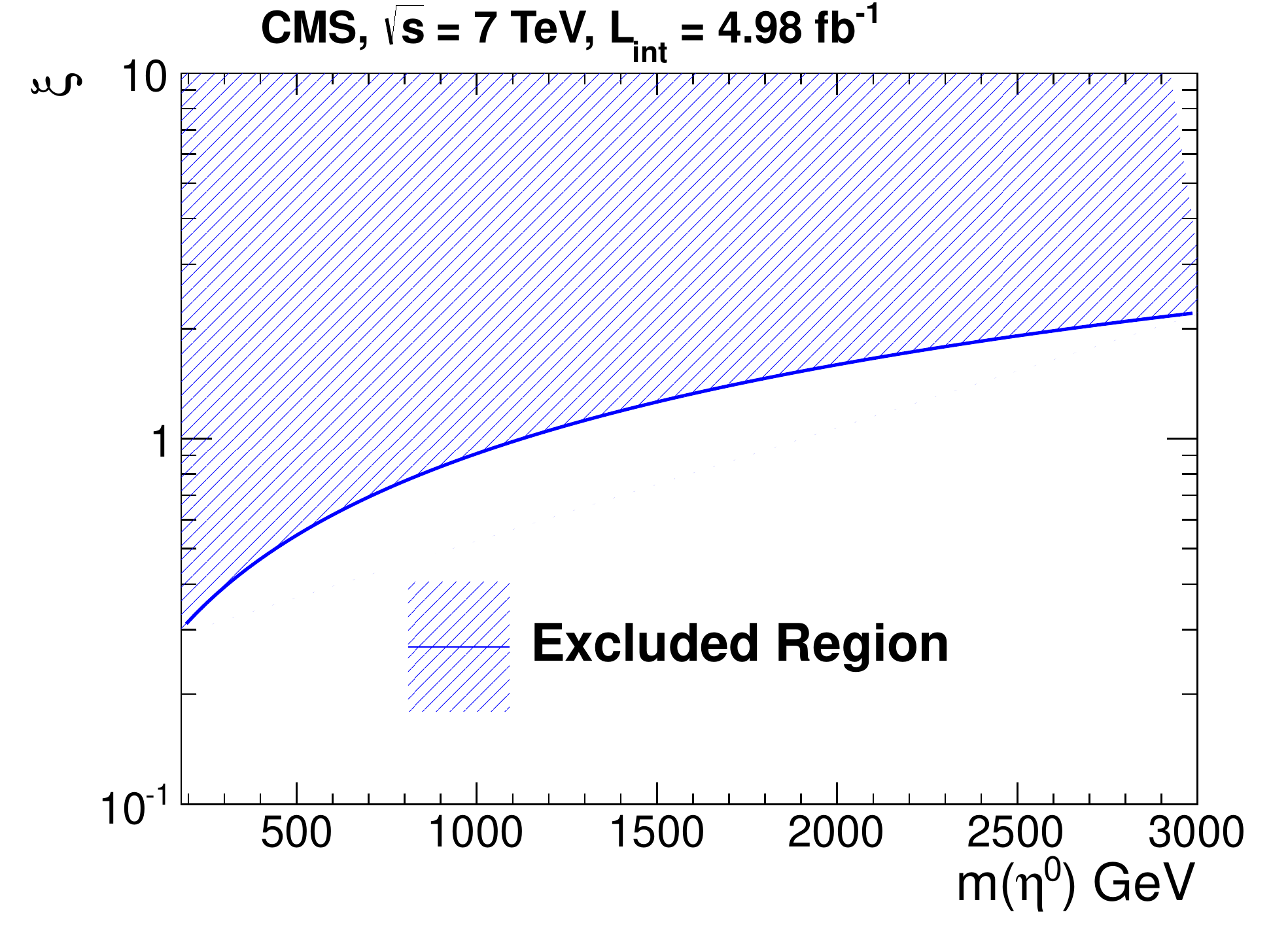}
\caption{Excluded regions in the parameter spaces of the
$\cPZpr$ (left) and MxFV models (right).  In the case of the $\zp$ model
we also show the $m(\zp)$ vs. $f_R$ region
consistent
with the
Tevatron $\ttbar$ forward-backward asymmetry measurements~\cite{fcnczprime}.
\label{fig:tt}}
\end{center}
\end{figure}

The limits on the parameter spaces of the $\zp$ and MxFV models
are shown in Fig.~\ref{fig:tt}.  These limits are based
on the lowest order cross section calculation.
Our bounds disfavor the $\zp$
model as an explanation of the Tevatron $\ttbar$
forward-backward asymmetry; the MxFV limits are significantly
more stringent than those of the CDF experiment~\cite{mxflv3}.

\subsection{Models with four top quarks and two LSPs from gluino pair production and decay via real or virtual top squarks}
\label{sec:stop}

\begin{figure}[htb]
\begin{center}
\includegraphics[height=0.33\linewidth]{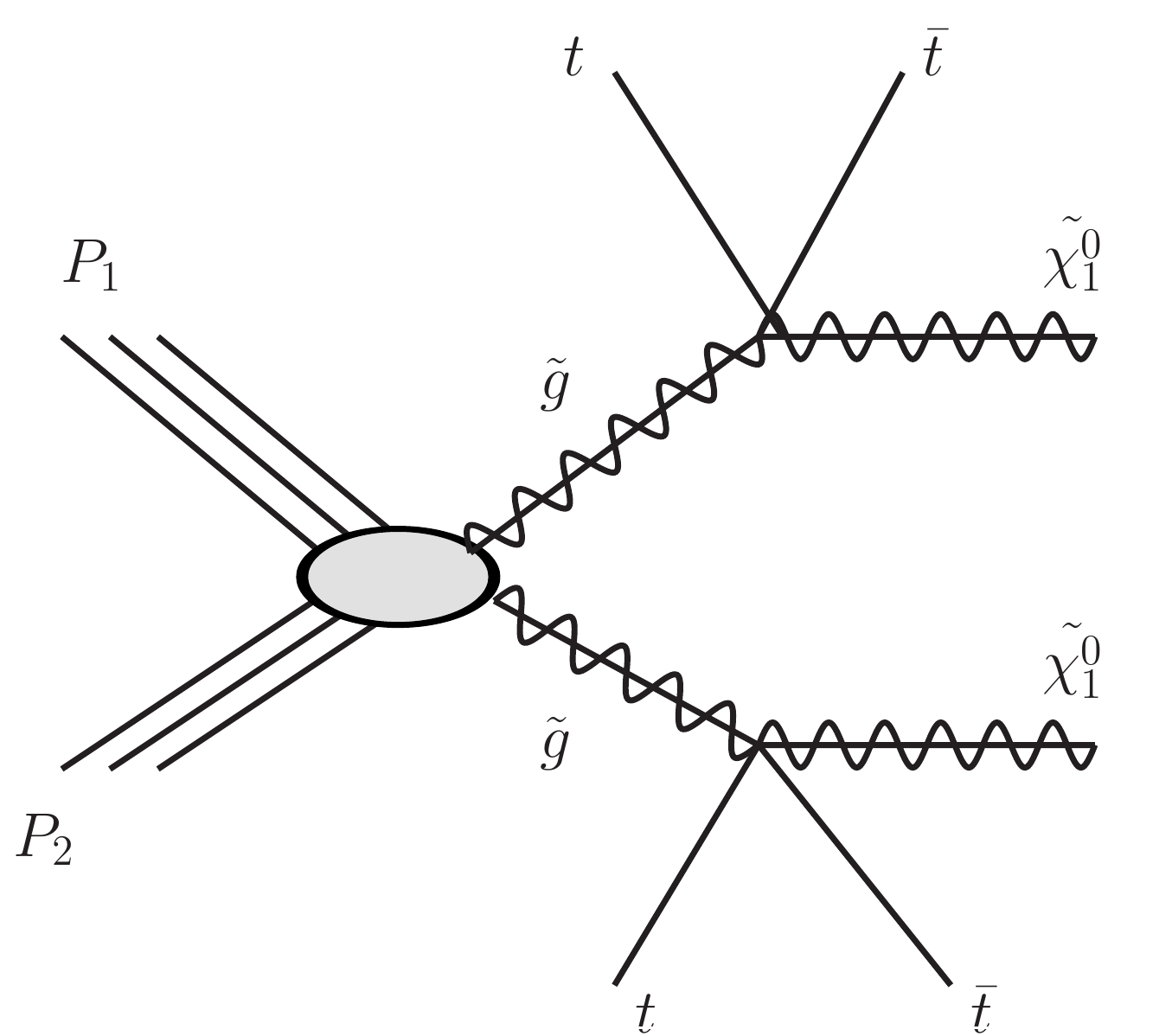}
\hspace{2 cm}
\includegraphics[height=0.33\linewidth]{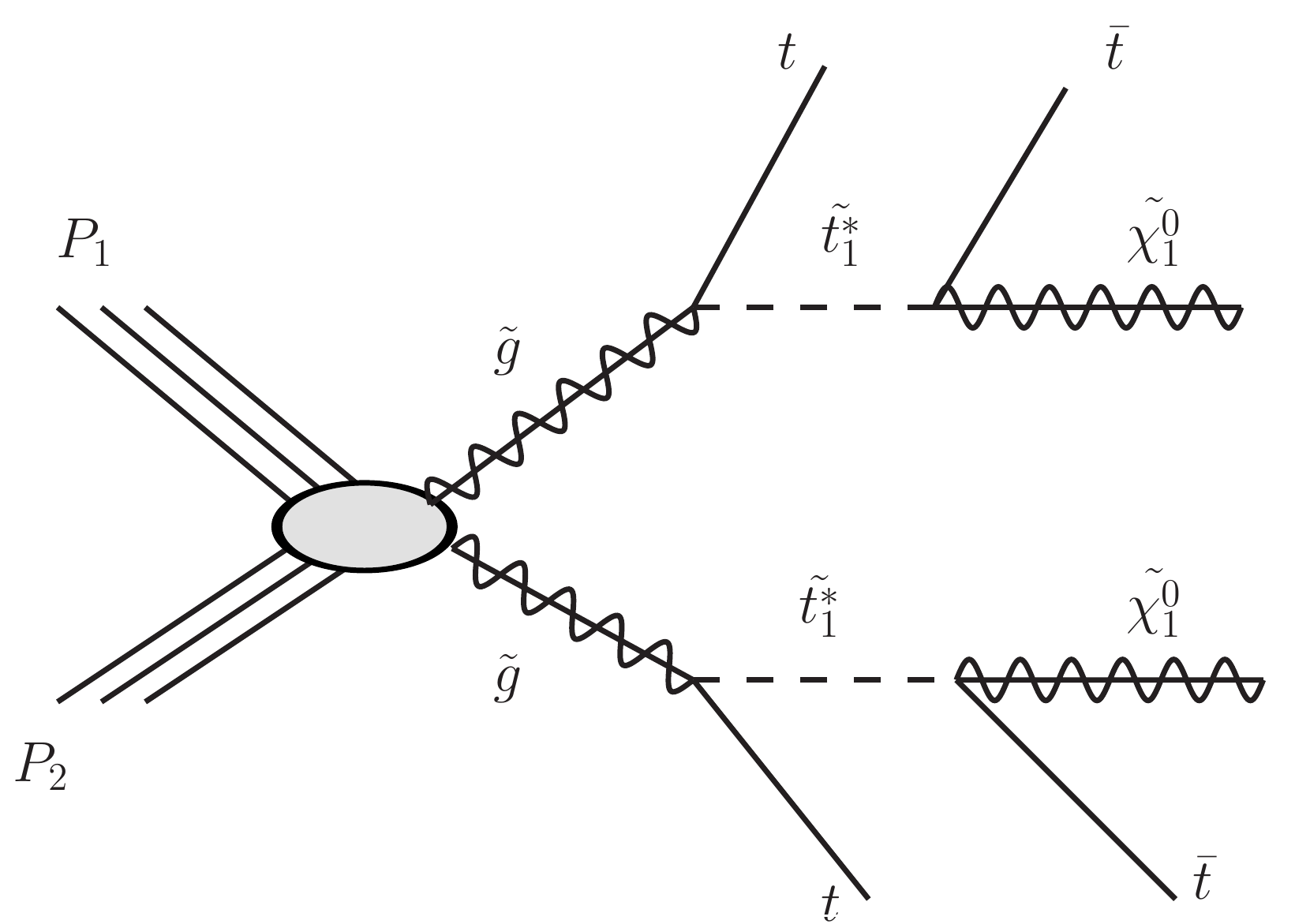}
\caption{Diagrams for models A1 (left) and
A2 (right).
\label{fig:stopFD}}
\end{center}
\end{figure}

In this Section we consider two SUSY models of gluino pair
production ($\pp \to \sGlu \sGlu$)
with top squarks
playing a dominant role in the decay of the gluino.  The gluino
decays under consideration are (see Fig.~\ref{fig:stopFD}):
\begin{itemize}
\item Model A1, three-body gluino decay mediated
by virtual stop: $\sGlu \to \PQt \PAQt \chiz_1$~\cite{stopVirtual,stopVirtualPRD,T1tttt};
\item Model A2,
two-body gluino decay to a top-stop pair:
$\sGlu \to \sTop_1 \PAQt$, $\sTop_1 \to \PQt \chiz_1$~\cite{wacker,naturalness4}.
\end{itemize}
The assumption of model A1 is that the gluino is lighter than all the squarks,
and that the stop is the lightest squark.
The dominant gluino decay channel
would then be
$\sGlu \to \PQt \PAQt \chiz_1$,
mediated by virtual top squarks.
Model A2
is the same as model
A1 but with top squarks light enough to be on-shell.  Both models
result in
$\PQt \PQt \PAQt \PAQt \chiz_1 \chiz_1$
final states, \ie,
final states with as many as four isolated high-$\pt$ leptons, four
\PQb quarks, several light-quark jets, and significant missing transverse energy
from the neutrinos in \PW\ decay and the LSPs.
For Model A1, the parameters are the gluino mass, $m(\sGlu)$,
and the LSP mass, $m(\chiz_1)$.  Model A2
has the
stop mass, $m(\sTop_1)$, as an
additional parameter.

These models are particularly interesting because naturalness arguments
suggest that the top squark should be relatively light.  A possible SUSY
scenario consistent with the initial data from the LHC
consists of a light stop,
with all other
squarks having evaded detection due to their very high mass.
Furthermore,
in order to preserve naturalness, the gluino cannot be too heavy either.
Thus, the possibility of a relatively light gluino decaying
predominantly into real
or virtual top squarks is very attractive; see Ref.~\cite{naturalness4}
for a recent discussion.

Signal events for models A1 and A2
are generated with \PYTHIA.
We find that for a large range of parameter space the most sensitive
signal region is SR6.  This is because these new physics
scenarios result in many jets and significant \MET.  Near the kinematic
boundaries, where the $\chiz_1$ has low momentum, SR4 and SR5
tend to be the most sensitive.

\begin{figure}[htb]
\begin{center}
\includegraphics[width=0.49\linewidth]{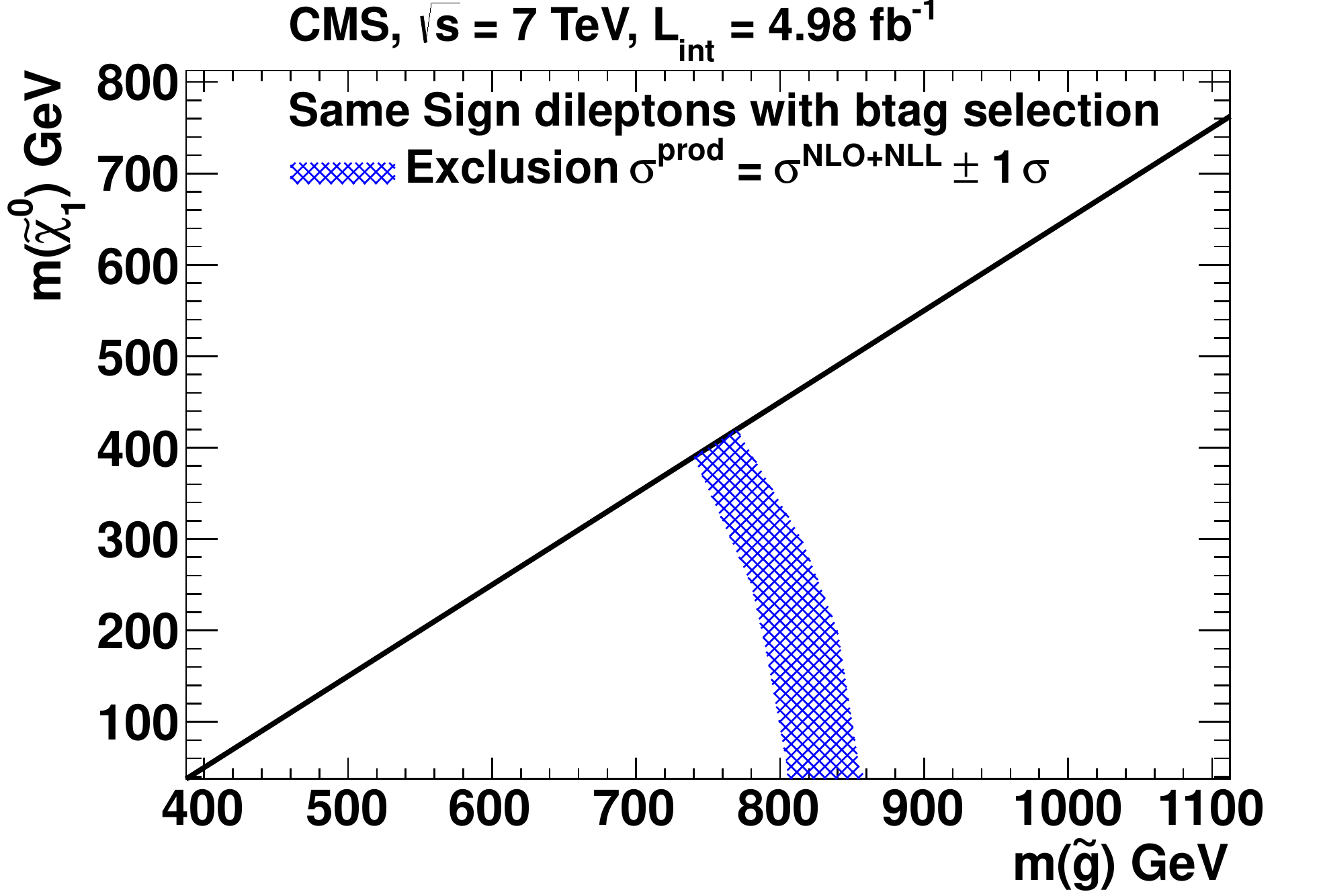}
\includegraphics[width=0.49\linewidth]{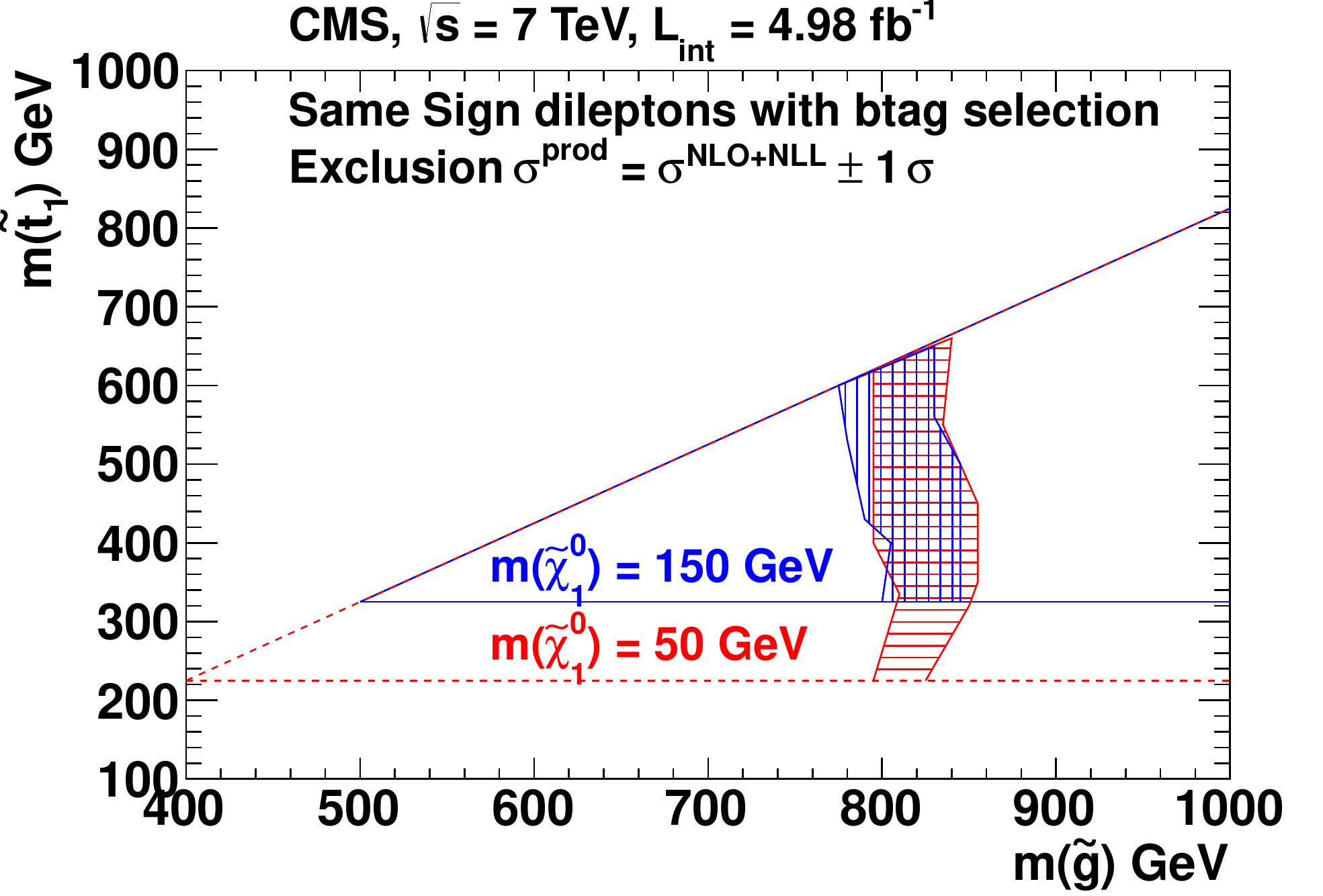}
\caption{Left plot: exclusion (95 \% CL) in the
$m(\chiz_1)-m(\sGlu)$
plane for model A1 (gluino decay via virtual top squarks).
Right plot: exclusion (95\% CL) in the
$m(\sTop_1)-m(\sGlu)$
plane for model A2
(gluino decay to on-shell top squarks).
The lines represent the kinematic boundaries of the models.
The regions to the left of the bands, and within the kinematic boundaries,
are excluded; the thicknesses of the bands represent the theoretical
uncertainties on the gluino pair production cross section from scale
and parton distribution functions (pdf) variations.
In the case of model A2
we show results for $m(\chiz_1)=50$\GeV (red, with dashed lines for the
kinematic boundaries) and $m(\chiz_1)=150$\GeV (blue, with solid line
for the kinematic boundary).
\label{fig:stop}}
\end{center}
\end{figure}

The limits on the parameter space of the A1 and
A2
models are displayed
in Fig.~\ref{fig:stop}.  These limits are based on the
next-to-leading-order (NLO)
and next-to-leading-log (NLL)
calculations of the gluino pair
production cross section~\cite{Kramer:2012bx,Kulesza:2008jb,Beenakker:2010nq}.

\subsection{Models with multiple top quarks and W-bosons from decays of bottom squarks}
\label{sec:sbottom}

\begin{figure}[htb]
\begin{center}
\includegraphics[width=0.49\linewidth]{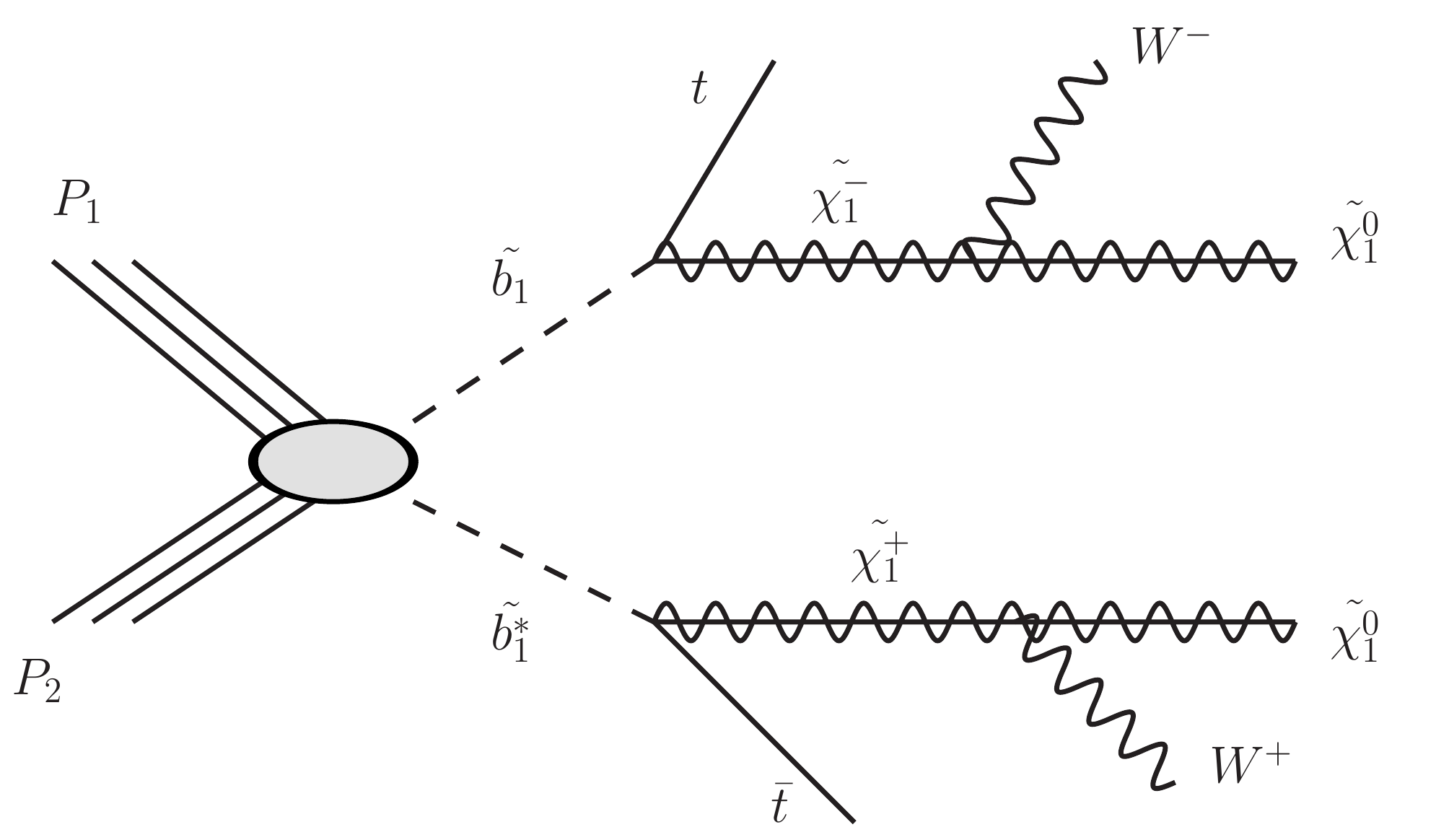}
\includegraphics[width=0.49\linewidth]{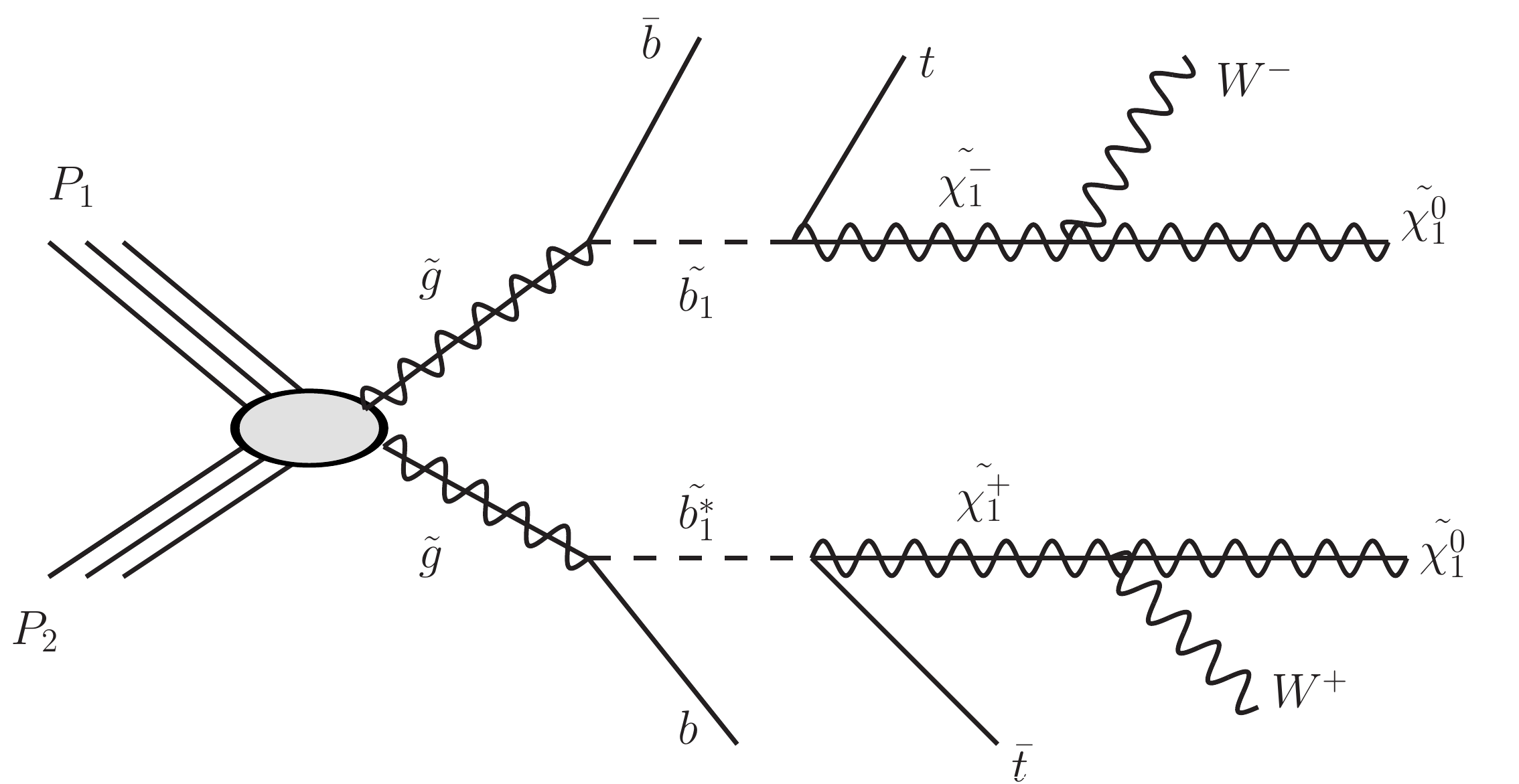}
\caption{Diagrams for models B1 (left) and
B2 (right).
\label{fig:sbFD}}
\end{center}
\end{figure}

\begin{figure}[bht]
\begin{center}
\includegraphics[width=0.49\linewidth]{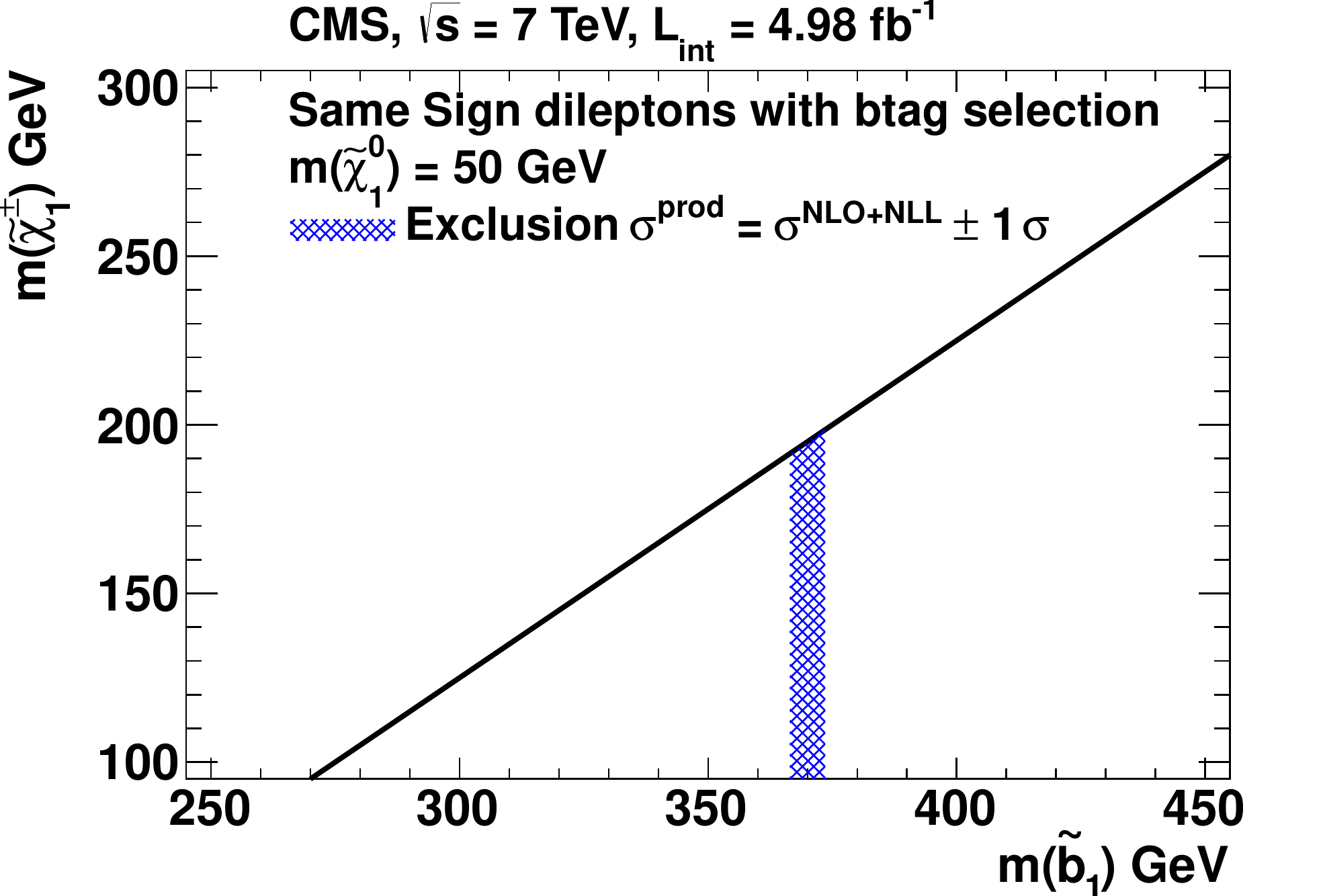}
\includegraphics[width=0.49\linewidth]{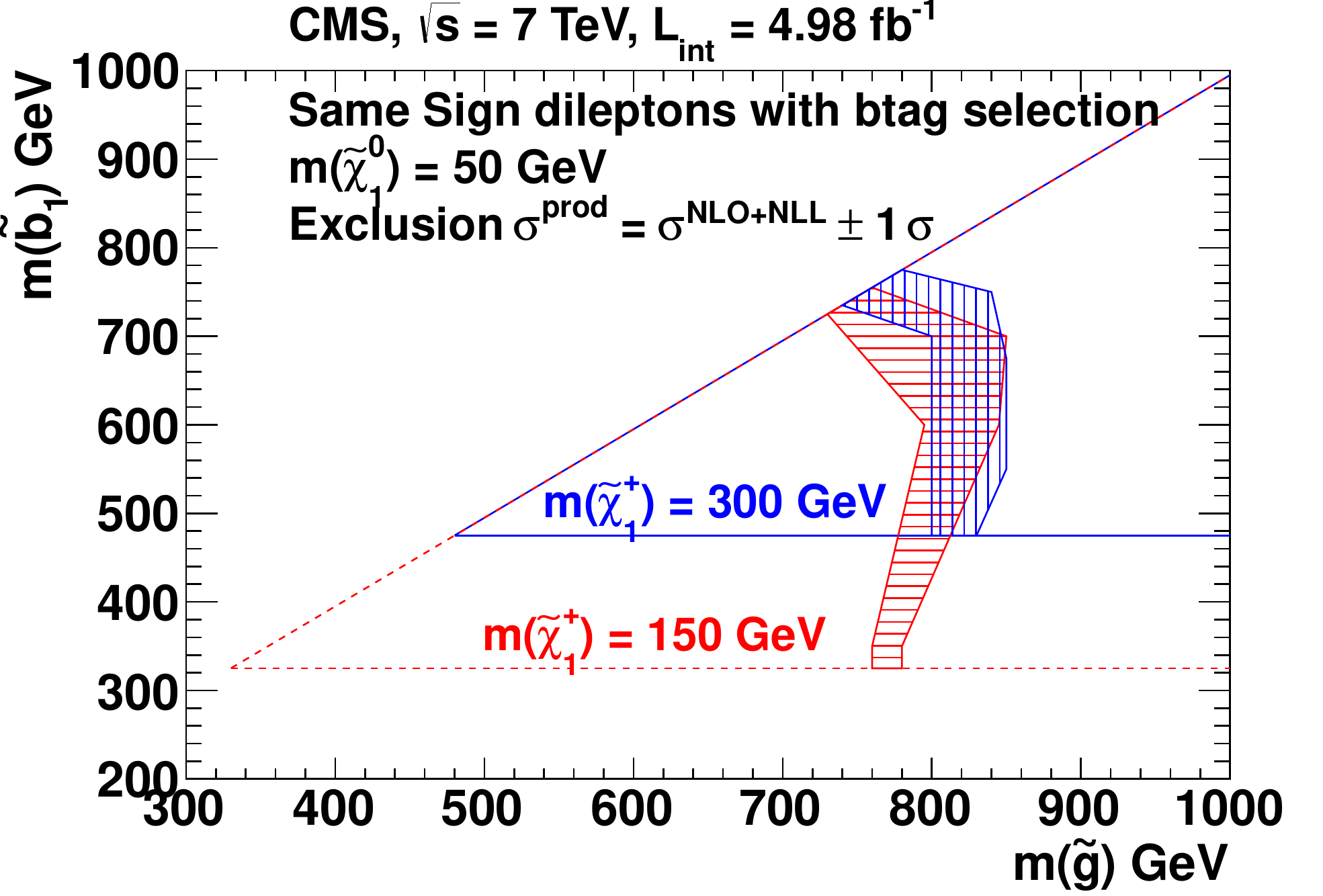}
\end{center}
\caption{Left plot: exclusion (95\% CL) in the
$m(\chipm_1) - m(\sBot_1)$
plane for model B1 (sbottom pair production);
Right plot: exclusion (95\% CL) in the $m(\sBot_1)-m(\sGlu)$
plane for model B2 (sbottom production from gluino decay).
The lines represent the kinematic boundaries of the models.
The regions to the left of the bands, and within the kinematic boundaries,
are excluded; the thicknesses of the bands represent the theoretical
uncertainties on the gluino and sbottom pair production cross section from scale
and parton distribution functions (pdf) variations.
In the case of model B2
we show results for $m(\chipm_1)=150$ \GeV (red, with dashed
line for the kinematic boundary) and $m(\chipm_1)=300$\GeV (blue, with
solid line for the kinematic boundary).
\label{fig:sbottom}}
\end{figure}

Here we study
possible SUSY signals with pairs of
bottom squarks decaying as
$\sBot_1 \to \PQt \chim_1$ and $\chim_1 \to \PWm \chiz_1$.
The production mechanisms are (see Fig.~\ref{fig:sbFD}):
\begin{itemize}
\item Model B1, sbottom pair production:  $\pp \to \sBot_1 \sBot_1^*$
\item Model B2, sbottom from gluino decay: $\pp \to \sGlu \sGlu$ or
or $\pp \to \sGlu \sBot_1$, followed by
$\sGlu \to \sBot_1 \PAQb$.
\end{itemize}

\noindent In scenarios where the sbottom is the lightest squark,
the gluino decay mode of model B2 would have the highest branching fraction .

The final states are then
$\PQt \PAQt \PWp \PWm \chiz_1 \chiz_1$
for model B1
and, for model B2, a mixture of
$\PQt \PQt \PWm \PWm$, $\PQt \PAQt \PWm \PWp$, and $\PAQt \PAQt \PWp \PWp$,
all with two $\chiz_1$ and two \PQb quarks.
For simplicity we consider only mass parameters
where the chargino and the \PW\ from chargino decay are on shell,
except for model B1, where the \PW\ is allowed to be off shell.

\begin{figure}[htb]
\begin{center}
\includegraphics[width=0.48\linewidth]{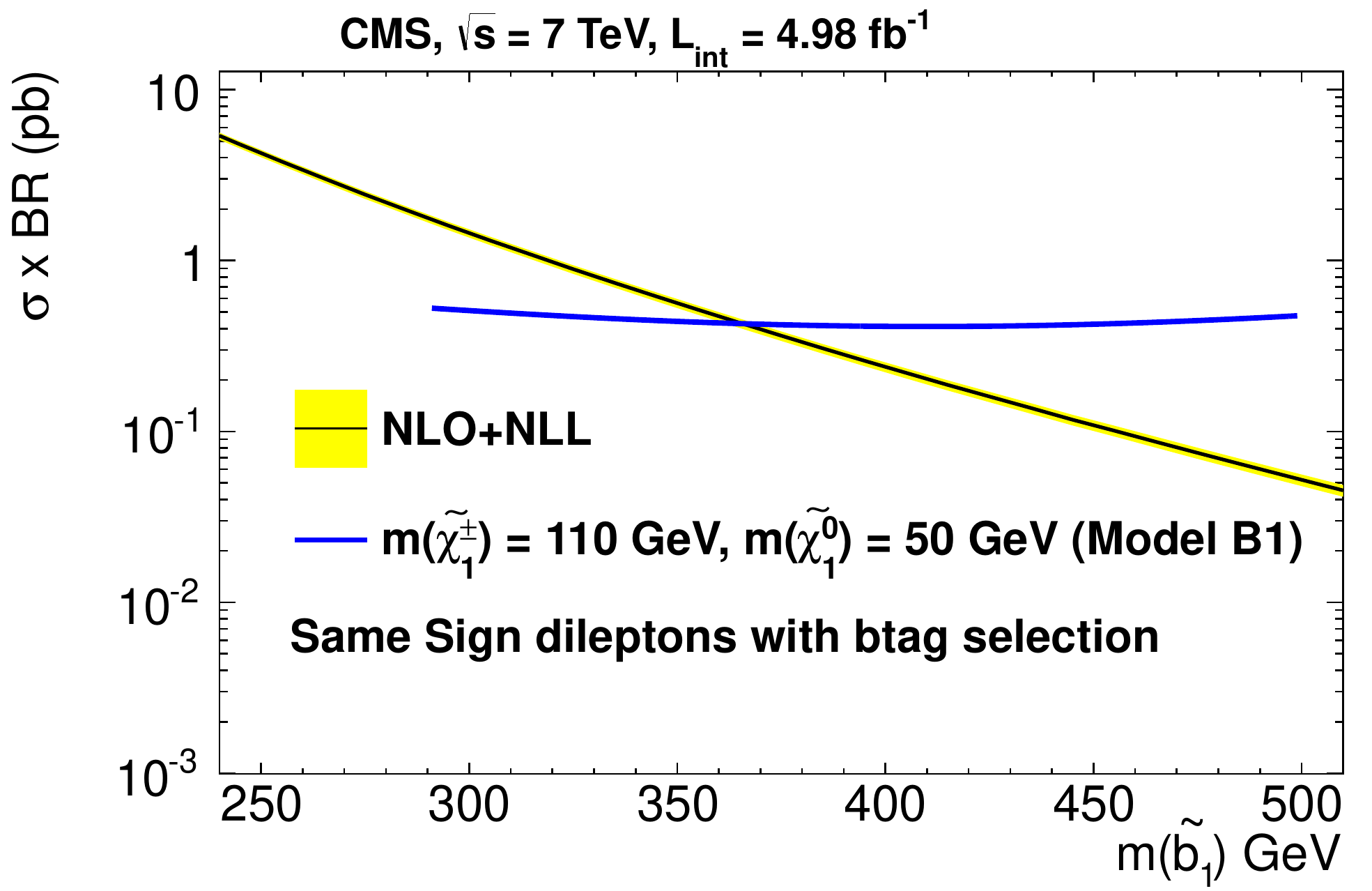}
\includegraphics[width=0.48\linewidth]{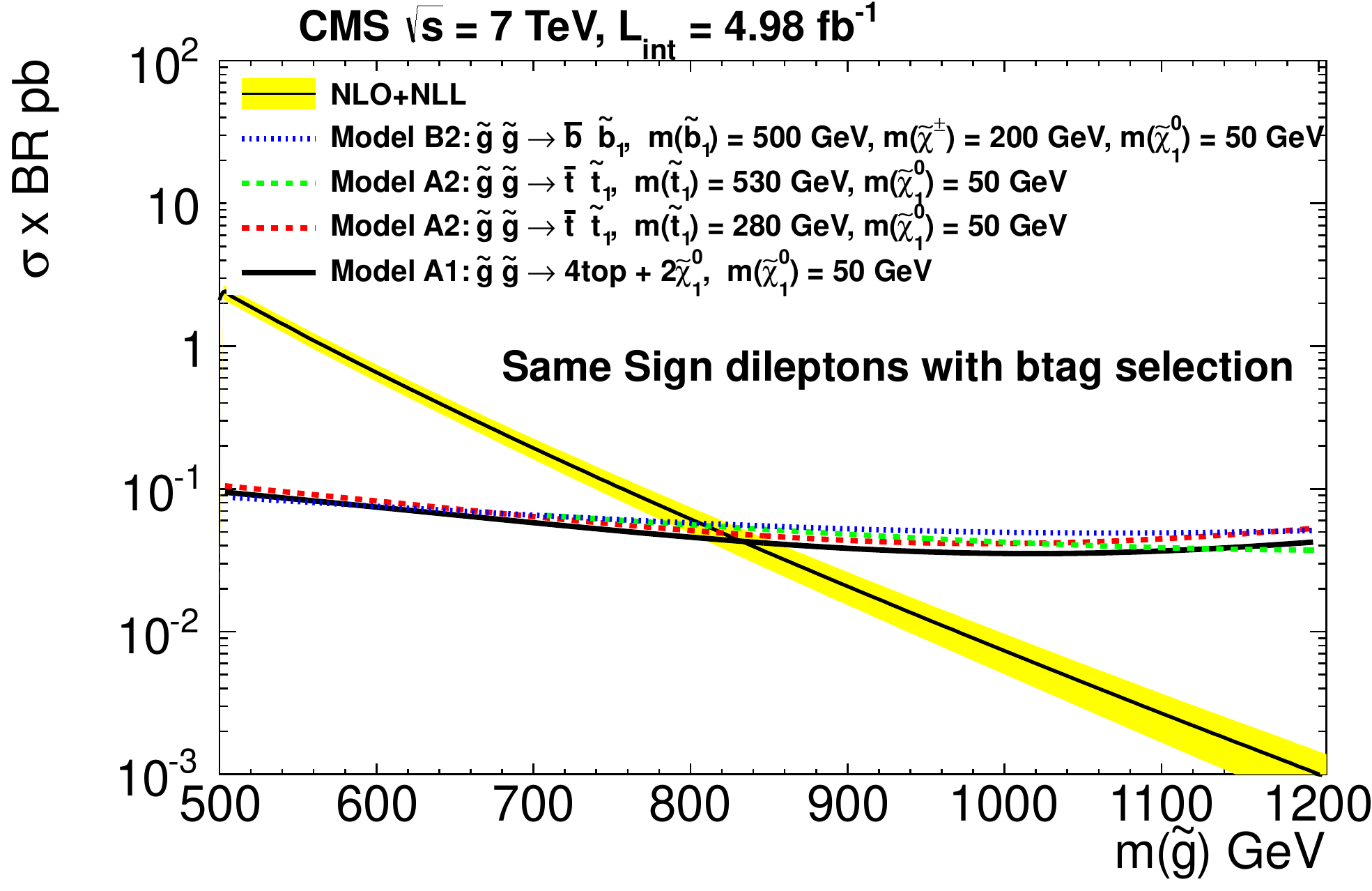}
\caption{Left plot: limits on the sbottom pair production
cross section compared with its expected value (NLO$+$NLL)
as a function
of sbottom mass
in model B1.  The cross section limit is insensitive to the
choice of LSP mass within the allowed kinematic range.
Right plot: limits on the gluino pair production cross section,
for models A1, A2, and B2,
compared with its expected value (NLO$+$NLL),
as a function of gluino mass.
\label{fig:sbottomLimit1d}}
\end{center}
\end{figure}

These final states yield up to four isolated high $\pt$ leptons, and
between
two and four bottom quarks.   For model B1 the parameters are the mass of
the sbottom, $m(\sBot_1)$, the mass of the chargino,
$m(\chipm_1)$, and the mass of the LSP, $m(\chiz_1)$. Model B2
has $m(\sGlu)$ as an additional parameter.

Signal events for models B1 and B2 were also generated with \PYTHIA.
The most sensitive signal regions
are SR1 and SR4 for model B1, and SR5 and SR6
for model B2.  The exclusion regions in parameter space are shown
in Fig.~\ref{fig:sbottom} and are based on the NLO+NLL calculations of the
production cross sections.

In Fig.~\ref{fig:sbottomLimit1d} (left)
we show the limits on the sbottom pair-production cross section
from model B1 together with expectations for this quantity.
The error band on the cross section curve reflects
the uncertainty in the choice of scale as well as the
associated pdf uncertainties.
Within the allowed kinematic range,
we exclude $m(\sBot_1)$ below 370\GeV for model B1.
The
limits on $\sigma(\pp \to \sGlu \sGlu)$ for
a few choices of the parameters of
A1, A2, and B2 are displayed in Fig.~\ref{fig:sbottomLimit1d} (right).
When compared with the expected gluino pair
production coss-section, we find that
the gluino mass limit is fairly insensitive
to the details of the decay chain, since the limit
is driven by the gluino cross section.
Models A1, A2, and B2 were also addressed in searches by the 
Atlas collaboration~\cite{ATLAS:2012ai,ATLAS:2012ah}.

\section{Conclusions}
\label{sec:conclusion}
We have presented results of a search for
same-sign dileptons with \PQb jets using the
CMS detector at the LHC based on
a \intLumi~data sample of \pp\ collisions at $\sqrt{s}$ = 7\TeV.
No significant deviations from the SM expectations are observed.

The data are used to set 95\% CL upper limits on the
number of new physics events for a number of plausible signal regions
defined in terms of requirements in \MET and $\HT$, the number of
\PQb-tagged jets (2 or 3), and also the sign of the leptons
(only positive dileptons
or both positive and negative dileptons).

We use these results to set a limit $\sigma(\pp \to \PQt \PQt) < 0.61$\unit{pb}
at 95\% CL, and to put bounds on
the parameter space of two models of
same-sign top pair production.
We also set limits on  two models of gluino decay into
on-shell or off-shell
top squarks, a model of sbottom pair production, and a model
of sbottom production from gluino decay. In addition,
we provide information to
interpret our limits in other models of new physics.

\section*{Acknowledgements}
We thank Johan Alwall, Ed Berger, Qing-Hong Cao, Chuan-Ren Chen,
Chong-Sheng Li, Hao Zhang, and Felix Yu
for discussions and help in implementing the \zp and MxFV models in \MADGRAPH.
We wish to congratulate our colleagues in the CERN accelerator departments for the excellent performance of the LHC machine.
We thank the technical and administrative staff at CERN and other CMS institutes, and acknowledge support from:
FMSR (Austria); FNRS and FWO (Belgium); CNPq, CAPES, FAPERJ, and FAPESP (Brazil);
MES (Bulgaria); CERN; CAS, MoST, and NSFC (China); COLCIENCIAS (Colombia);
MSES (Croatia); RPF (Cyprus); Academy of Sciences and NICPB (Estonia); Academy of Finland, MEC, and HIP (Finland);
CEA and CNRS/IN2P3 (France); BMBF, DFG, and HGF (Germany); GSRT (Greece); OTKA and NKTH (Hungary); DAE and DST (India);
IPM (Iran); SFI (Ireland); INFN (Italy); NRF and WCU (Korea); LAS (Lithuania);
CINVESTAV, CONACYT, SEP, and UASLP-FAI (Mexico); MSI (New Zealand); PAEC (Pakistan);
MSHE and NSC (Poland); FCT (Portugal); JINR (Armenia, Belarus, Georgia, Ukraine, Uzbekistan);
MON, RosAtom, RAS and RFBR (Russia); MSTD (Serbia); MICINN and CPAN (Spain);
Swiss Funding Agencies (Switzerland); NSC (Taipei); TUBITAK and TAEK (Turkey); STFC (United Kingdom); DOE and NSF (USA).

\bibliography{auto_generated}   

\providecommand{\href}[2]{#2}\begingroup\raggedright\begin{thebibliography}{10}%
\makeatletter
\providecommand{\hrefCMSnoop }[0]{\@secondoftwo}%
\makeatother
\providecommand{\doi}{\texttt{doi:}\begingroup \urlstyle{tt}\Url}

\bibitem{naturalness1}
\hrefCMSnoop {} {A.~G. Cohen, D.~B. Kaplan, and A.~E. Nelson, ``{The more
  Minimal Supersymmetric Standard Model}'',} \textit{ Phys. Lett. B} \textbf{
  388} (1996) 588,
  \href{http://dx.doi.org/10.1016/S0370-2693(96)01183-5}{\doi{10.1016/S0370-2693(96)01183-5}},
\href{http://www.arXiv.org/abs/hep-ph/9607394}{\texttt{ arXiv:hep-ph/9607394}}.

\bibitem{naturalness2}
\hrefCMSnoop {} {S.~Dimopoulos and G.~F. Giudice, ``{Naturalness constraints in
  supersymmetric theories with non-universal soft terms}'',} \textit{ Phys.
  Lett. B} \textbf{ 357} (1995) 573,
  \href{http://dx.doi.org/10.1016/0370-2693(95)00961-J}{\doi{10.1016/0370-2693(95)00961-J}},
\href{http://www.arXiv.org/abs/hep-ph/9507282}{\texttt{ arXiv:hep-ph/9507282}}.

\bibitem{naturalness3}
\hrefCMSnoop {} {R.~Barbieri, G.~R. Dvali, and L.~J. Hall, ``{Predictions from
  a U(2) flavor symmetry in supersymmetric theories}'',} \textit{ Phys. Lett.
  B} \textbf{ 377} (1996) 76,
  \href{http://dx.doi.org/10.1016/0370-2693(96)00318-8}{\doi{10.1016/0370-2693(96)00318-8}},
\href{http://www.arXiv.org/abs/hep-ph/9512388}{\texttt{ arXiv:hep-ph/9512388}}.

\bibitem{naturalness4}
\hrefCMSnoop {} {M.~Papucci, J.~T. Ruderman, and A.~Weiler, ``{Natural SUSY
  Endures}'',} (2011).
\href{http://www.arXiv.org/abs/1110.6926}{\texttt{ arXiv:1110.6926}}.

\bibitem{Csaki:2012fh}
\hrefCMSnoop {} {C.~Csaki, L.~Randall, and J.~Terning, ``{Light Stops from
  Seiberg Duality}'',} (2012).
\href{http://www.arXiv.org/abs/1201.1293}{\texttt{ arXiv:1201.1293}}.

\bibitem{fcnczprime}
\hrefCMSnoop {} {E.~L. Berger, Q.-H. Cao, C.-R. Chen{ et~al.}, ``{Top Quark
  Forward-Backward Asymmetry and Same-Sign Top Quark Pairs}'',} \textit{ Phys.
  Rev. Lett.} \textbf{ 106} (2011) 201801,
  \href{http://dx.doi.org/10.1103/PhysRevLett.106.201801}{\doi{10.1103/PhysRevLett.106.201801}},
\href{http://www.arXiv.org/abs/1101.5625}{\texttt{ arXiv:1101.5625}}.

\bibitem{Buckley}
\hrefCMSnoop {} {M.~R. Buckley, D.~Hooper, J.~Kopp{ et~al.}, ``{Light Z' Bosons
  at the Tevatron}'',} \textit{ Phys. Rev. D} \textbf{ 83} (2011) 115013,
  \href{http://dx.doi.org/10.1103/PhysRevD.83.115013}{\doi{10.1103/PhysRevD.83.115013}},
\href{http://www.arXiv.org/abs/1103.6035}{\texttt{ arXiv:1103.6035}}.

\bibitem{d0:fwtop}
\hrefCMSnoop {} {{ D0} Collaboration, ``{Measurement of the Forward-Backward
  Charge Asymmetry in Top-Quark Pair Production}'',} \textit{ Phys. Rev. Lett.}
  \textbf{ 100} (2008) 142002,
  \href{http://dx.doi.org/10.1103/PhysRevLett.100.142002}{\doi{10.1103/PhysRevLett.100.142002}},
\href{http://www.arXiv.org/abs/0712.0851}{\texttt{ arXiv:0712.0851}}.

\bibitem{cdf:fwtop1}
\hrefCMSnoop {} {{ CDF} Collaboration, ``{Forward-Backward Asymmetry in
  Top-Quark Production in $p\bar{p}$ Collisions at $\sqrt{s}=1.96$ TeV}'',}
  \textit{ Phys. Rev. Lett.} \textbf{ 101} (2008) 202001,
  \href{http://dx.doi.org/10.1103/PhysRevLett.101.202001}{\doi{10.1103/PhysRevLett.101.202001}},
\href{http://www.arXiv.org/abs/0806.2472}{\texttt{ arXiv:0806.2472}}.

\bibitem{cdf:fwtop2}
\hrefCMSnoop {} {{ CDF} Collaboration, ``{Evidence for a mass dependent
  forward-backward asymmetry in top quark pair production}'',} \textit{ Phys.
  Rev. D} \textbf{ 83} (2011) 112003,
  \href{http://dx.doi.org/10.1103/PhysRevD.83.112003}{\doi{10.1103/PhysRevD.83.112003}},
\href{http://www.arXiv.org/abs/1101.0034}{\texttt{ arXiv:1101.0034}}.

\bibitem{mxflv1}
\hrefCMSnoop {} {S.~Bar-Shalom and A.~Rajaraman, ``{Models and phenomenology of
  maximal flavor violation}'',} \textit{ Phys. Rev. D} \textbf{ 77} (2008)
  095011,
  \href{http://dx.doi.org/10.1103/PhysRevD.77.095011}{\doi{10.1103/PhysRevD.77.095011}},
\href{http://www.arXiv.org/abs/0711.3193}{\texttt{ arXiv:0711.3193}}.

\bibitem{mxflv2}
\hrefCMSnoop {} {S.~Bar-Shalom, A.~Rajaraman, D.~Whiteson{ et~al.}, ``{Collider
  signals of maximal flavor violation: Same-sign leptons from same-sign top
  quarks at the Fermilab Tevatron}'',} \textit{ Phys. Rev. D} \textbf{ 78}
  (2008) 033003,
  \href{http://dx.doi.org/10.1103/PhysRevD.78.033003}{\doi{10.1103/PhysRevD.78.033003}},
\href{http://www.arXiv.org/abs/0803.3795}{\texttt{ arXiv:0803.3795}}.

\bibitem{mxflv3}
\hrefCMSnoop {} {{ CDF} Collaboration, ``{Search for Maximal Flavor Violating
  Scalars in Same-Charge Lepton Pairs in $p \bar{p}$ Collisions at $\sqrt{s}$ =
  1.96 TeV}'',} \textit{ Phys. Rev. Lett.} \textbf{ 102} (2009) 041801,
  \href{http://dx.doi.org/10.1103/PhysRevLett.102.041801}{\doi{10.1103/PhysRevLett.102.041801}},
\href{http://www.arXiv.org/abs/0809.4903}{\texttt{ arXiv:0809.4903}}.

\bibitem{sspaper2010}
\hrefCMSnoop {} {{ CMS} Collaboration, ``{Search for new physics with same-sign
  isolated dilepton events with jets and missing transverse energy at the
  LHC}'',} \textit{ JHEP} \textbf{ 1106} (2011) 077,
  \href{http://dx.doi.org/10.1007/JHEP06(2011)077}{\doi{10.1007/JHEP06(2011)077}},
\href{http://www.arXiv.org/abs/1104.3168}{\texttt{ arXiv:1104.3168}}.

\bibitem{sspaper2011}
\hrefCMSnoop {} {{ CMS} Collaboration, ``{Search for new physics with same-sign
  isolated dilepton events with jets and missing energy}'',} (2012).
\href{http://www.arXiv.org/abs/1205.6615}{\texttt{ arXiv:1205.6615}}.

\bibitem{JINST}
\hrefCMSnoop {} {{ CMS} Collaboration, ``The {CMS} experiment at the {CERN}
  {LHC}'',} \textit{ JINST} \textbf{ 3} (2008) S08004,
\href{http://dx.doi.org/10.1088/1748-0221/3/08/S08004}{\doi{10.1088/1748-0221/3/08/S08004}}.

\bibitem{sstop}
\hrefCMSnoop {} {{ CMS} Collaboration, ``{Search for same-sign top-quark pair
  production at $\sqrt{s}$ = 7 TeV and limits on flavour changing neutral
  currents in the top sector}'',} \textit{ JHEP} \textbf{ 1108} (2011) 005,
  \href{http://dx.doi.org/10.1007/JHEP08(2011)005}{\doi{10.1007/JHEP08(2011)005}},
\href{http://www.arXiv.org/abs/1106.2142}{\texttt{ arXiv:1106.2142}}.

\bibitem{EGMPAS}
\href {http://cdsweb.cern.ch/record/1299116} {{ CMS} Collaboration, ``Electron
  Reconstruction and Identification at $\sqrt{s} = 7$ {TeV}'',} CMS Physics
  Analysis Summary CMS-PAS-EGM-10-004, (2010).

\bibitem{MUOPAS}
\href {http://cdsweb.cern.ch/record/1279140} {{ CMS} Collaboration,
  ``Performance of muon identification in pp collisions at $\sqrt{s}$ = 7
  {TeV}'',} CMS Physics Analysis Summary CMS-PAS-MUO-10-002, (2010).

\bibitem{PFPAS}
\href {http://cdsweb.cern.ch/record/1194487} {{ CMS} Collaboration,
  ``Particle--Flow Event Reconstruction in {CMS} and Performance for Jets,
  Taus, and {\MET}'',} CMS Physics Analysis Summary CMS-PAS-PFT-09-001, (2009).

\bibitem{JES}
\hrefCMSnoop {} {{ CMS} Collaboration, ``{Determination of Jet Energy
  Calibration and Transverse Momentum Resolution in CMS}'',} \textit{ JINST}
  \textbf{ 6} (2011) P11002,
  \href{http://dx.doi.org/10.1088/1748-0221/6/11/P11002}{\doi{10.1088/1748-0221/6/11/P11002}},
\href{http://www.arXiv.org/abs/1107.4277}{\texttt{ arXiv:1107.4277}}.

\bibitem{MET}
\hrefCMSnoop {} {{ CMS} Collaboration, ``{Missing transverse energy performance
  of the CMS detector}'',} \textit{ JINST} \textbf{ 6} (2011) P09001,
  \href{http://dx.doi.org/10.1088/1748-0221/6/09/P09001}{\doi{10.1088/1748-0221/6/09/P09001}},
\href{http://www.arXiv.org/abs/1106.5048}{\texttt{ arXiv:1106.5048}}.

\bibitem{Cacciari:2008gp}
\hrefCMSnoop {} {M.~Cacciari, G.~P. Salam, and G.~Soyez, ``{The anti-$k_t$ jet
  clustering algorithm}'',} \textit{ JHEP} \textbf{ 0804} (2008) 063,
  \href{http://dx.doi.org/10.1088/1126-6708/2008/04/063}{\doi{10.1088/1126-6708/2008/04/063}},
\href{http://www.arXiv.org/abs/0802.1189}{\texttt{ arXiv:0802.1189}}.

\bibitem{CMS-PAS-BTV-11-002}
\href {http://cdsweb.cern.ch/record/1395489} {{ CMS} Collaboration, ``Status of
  b-tagging tools for 2011 data analysis'',} CMS Physics Analysis Summary
  CMS-PAS-BTV-11-002, (2011).

\bibitem{CMS-PAS-BTV-11-003}
\href {http://cdsweb.cern.ch/record/1421611} {{ CMS} Collaboration,
  ``Measurement of the b-tagging efficiency using \ttbar\ events'',} CMS
  Physics Analysis Summary CMS-PAS-BTV-11-003, (2011).

\bibitem{MADGRAPH4}
\hrefCMSnoop {} {J.~Alwall, P.~Demin, S.~de~Visscher{ et~al.},
  ``{MadGraph/MadEvent v4: the new web generation}'',} \textit{ JHEP} \textbf{
  0709} (2007) 028,
  \href{http://dx.doi.org/10.1088/1126-6708/2007/09/028}{\doi{10.1088/1126-6708/2007/09/028}},
\href{http://www.arXiv.org/abs/0706.2334}{\texttt{ arXiv:0706.2334}}.

\bibitem{PYTHIA}
\hrefCMSnoop {} {T.~Sjostrand, S.~Mrenna, and P.~Z. Skands, ``{PYTHIA 6.4
  Physics and Manual}'',} \textit{ JHEP} \textbf{ 0605} (2006) 026,
  \href{http://dx.doi.org/10.1088/1126-6708/2006/05/026}{\doi{10.1088/1126-6708/2006/05/026}},
\href{http://www.arXiv.org/abs/hep-ph/0603175}{\texttt{ arXiv:hep-ph/0603175}}.

\bibitem{Campbell:2012dh}
\hrefCMSnoop {} {J.~M. Campbell and R.~K. Ellis, ``{$t\bar{t}W^\pm$ production
  and decay at NLO}'',} (2012).
\href{http://www.arXiv.org/abs/1204.5678}{\texttt{ arXiv:1204.5678}}.

\bibitem{ttzNLO}
\hrefCMSnoop {} {A.~Kardos, Z.~Trocsanyi, and C.~Papadopoulos, ``{Top quark
  pair production in association with a Z-boson at NLO accuracy}'',} \textit{
  Phys. Rev. D} \textbf{ 85} (2012) 054015,
  \href{http://dx.doi.org/10.1103/PhysRevD.85.054015}{\doi{10.1103/PhysRevD.85.054015}},
\href{http://www.arXiv.org/abs/1111.0610}{\texttt{ arXiv:1111.0610}}.

\bibitem{Garzelli:2011is}
\hrefCMSnoop {} {M.~V. Garzelli, A.~Kardos, C.~G. Papadopoulos{ et~al.},
  ``{Z$^0$-boson production in association with a $\ttbar$ pair at
  next-to-leading order accuracy with parton shower effects}'',} \textit{ Phys.
  Rev. D} \textbf{ 85} (2012) 074022,
  \href{http://dx.doi.org/10.1103/PhysRevD.85.074022}{\doi{10.1103/PhysRevD.85.074022}},
\href{http://www.arXiv.org/abs/1111.1444}{\texttt{ arXiv:1111.1444}}.

\bibitem{rpv}
\hrefCMSnoop {} {G.~R. Farrar and P.~Fayet, ``{Phenomenology of the production,
  decay, and detection of new hadronic states associated with
  supersymmetry}'',} \textit{ Phys. Lett. B} \textbf{ 76} (1978) 575,
\href{http://dx.doi.org/10.1016/0370-2693(78)90858-4}{\doi{10.1016/0370-2693(78)90858-4}}.

\bibitem{rpvnomet}
\hrefCMSnoop {} {P.~Fileviez~Perez and S.~Spinner, ``{The Minimal Theory for
  R-parity Violation at the LHC}'',} \textit{ JHEP} \textbf{ 1204} (2012) 118,
  \href{http://dx.doi.org/10.1007/JHEP04(2012)118}{\doi{10.1007/JHEP04(2012)118}},
\href{http://www.arXiv.org/abs/1201.5923}{\texttt{ arXiv:1201.5923}}.

\bibitem{rpvwmet}
\hrefCMSnoop {} {C.~Brust, A.~Katz, S.~Lawrence{ et~al.}, ``{SUSY, the Third
  Generation and the LHC}'',} \textit{ JHEP} \textbf{ 1203} (2012) 103,
  \href{http://dx.doi.org/10.1007/JHEP03(2012)103}{\doi{10.1007/JHEP03(2012)103}},
\href{http://www.arXiv.org/abs/1110.6670}{\texttt{ arXiv:1110.6670}}.

\bibitem{Dreiner:2012mn}
\hrefCMSnoop {} {H.~K. Dreiner and T.~Stefaniak, ``{Bounds on R-parity
  Violation from Resonant Slepton Production at the LHC}'',} (2012).
\href{http://www.arXiv.org/abs/1201.5014}{\texttt{ arXiv:1201.5014}}.

\bibitem{Junk:1999kv}
\hrefCMSnoop {} {T.~Junk, ``{Confidence level computation for combining
  searches with small statistics}'',} \textit{ Nucl. Instrum. Meth. A} \textbf{
  434} (1999) 435,
  \href{http://dx.doi.org/10.1016/S0168-9002(99)00498-2}{\doi{10.1016/S0168-9002(99)00498-2}},
\href{http://www.arXiv.org/abs/hep-ex/9902006}{\texttt{ arXiv:hep-ex/9902006}}.

\bibitem{ATL-PHYS-PUB-2011-011}
\href {https://cdsweb.cern.ch/record/1379837} {{ATLAS and CMS Collaborations},
  ``Procedure for the LHC Higgs boson search combination in summer 2011'',}
  ATL-PHYS-PUB-2011-011, {CMS NOTE-2011/005}, (2011).

\bibitem{PTDR2}
\hrefCMSnoop {} {{CMS Collaboration}, ``{CMS} technical design report, volume
  {II}: {Physics} performance'',} \textit{ J. Phys. G} \textbf{ 34} (2007) 995,
\href{http://dx.doi.org/10.1088/0954-3899/34/6/S01}{\doi{10.1088/0954-3899/34/6/S01}}.

\bibitem{CMS-PAS-SMP-12-008}
\href {http://cdsweb.cern.ch/record/1434360} {{ CMS} Collaboration, ``Absolute
  Calibration of the Luminosity Measurement at {CMS}: {W}inter 2012 Update'',}
  CMS Physics Analysis Summary CMS-PAS-SMP-12-008, (2012).

\bibitem{Abdullin:1328345}
{ CMS} Collaboration, \hrefCMSnoop {} {S.~Abdullin, P.~Azzi, F.~Beaudette{
  et~al.}, ``Fast simulation of the {CMS} detector at {LHC}'',} in \textit{
  International Conference on Computing in High Energy and Nuclear Physics
  (CHEP 2010)}.
\newblock IOP, 2011.
\newblock Journal of Physics: Conference Series.
  \href{http://dx.doi.org/10.1088/1742-6596/331/3/032049}{\doi{10.1088/1742-6596/331/3/032049}}.

\bibitem{CMS-DP-2010-039}
\href {http://cdsweb.cern.ch/record/1309890} {{ CMS} Collaboration,
  ``Comparison of the Fast Simulation of CMS with the first LHC data'',} CMS
  Detector Performance Summary CMS-DP-2010-039, (2010).

\bibitem{cmsdil}
\hrefCMSnoop {} {{ CMS} Collaboration, ``{Measurement of the $\ttbar$
  production cross section and the top quark mass in the dilepton channel in pp
  collisions at $\sqrt{s}$ =7 TeV}'',} \textit{ JHEP} \textbf{ 1107} (2011)
  049,
  \href{http://dx.doi.org/10.1007/JHEP07(2011)049}{\doi{10.1007/JHEP07(2011)049}},
\href{http://www.arXiv.org/abs/1105.5661}{\texttt{ arXiv:1105.5661}}.

\bibitem{Khachatryan:2010ez}
\hrefCMSnoop {} {{ CMS} Collaboration, ``{First measurement of the cross
  section for top-quark pair production in proton-proton collisions at
  $\sqrt{s}=7$ TeV}'',} \textit{ Phys. Lett. B} \textbf{ 695} (2011) 424,
  \href{http://dx.doi.org/10.1016/j.physletb.2010.11.058}{\doi{10.1016/j.physletb.2010.11.058}},
\href{http://www.arXiv.org/abs/1010.5994}{\texttt{ arXiv:1010.5994}}.

\bibitem{sstopatlas}
\hrefCMSnoop {} {{ ATLAS} Collaboration, ``{Search for same-sign top-quark
  production and fourth-generation down-type quarks in pp collisions at sqrt(s)
  = 7 TeV with the ATLAS detector}'',} \textit{ JHEP} \textbf{ 1204} (2012)
  069,
  \href{http://dx.doi.org/10.1007/JHEP04(2012)069}{\doi{10.1007/JHEP04(2012)069}},
\href{http://www.arXiv.org/abs/1202.5520}{\texttt{ arXiv:1202.5520}}.

\bibitem{cdfth2}
\hrefCMSnoop {} {J.~A. Aguilar-Saavedra, ``{Effective four-fermion operators in
  top physics: A Roadmap}'',} \textit{ Nucl. Phys. B} \textbf{ 843} (2011) 638,
  \href{http://dx.doi.org/10.1016/j.nuclphysb.2010.10.015}{\doi{10.1016/j.nuclphysb.2010.10.015}},
\href{http://www.arXiv.org/abs/1008.3562}{\texttt{ arXiv:1008.3562}}.

\bibitem{cdflimit}
\href
  {http://www-cdf.fnal.gov/physics/exotic/r2a/20110407.samesigndileptons/sstops.pdf}
  {{ {CDF}} Collaboration, ``Search for like-sign top quark pair production at
  CDF with 6.1 fb$^{-1}$'',} CDF Public Note {CDF-PHYS-EXO-PUBLIC-10466},
  (2011).

\bibitem{Meade:2007js}
\hrefCMSnoop {} {P.~Meade and M.~Reece, ``{BRIDGE: Branching Ratio
  Inquiry/Decay Generated Events}'',} (2007).
\href{http://www.arXiv.org/abs/hep-ph/0703031}{\texttt{ arXiv:hep-ph/0703031}}.

\bibitem{stopVirtual}
\hrefCMSnoop {} {B.~S. Acharya, P.~Grajek, G.~L. Kane{ et~al.}, ``{Identifying
  Multi-Top Events from Gluino Decay at the LHC}'',} (2009).
\href{http://www.arXiv.org/abs/0901.3367}{\texttt{ arXiv:0901.3367}}.

\bibitem{stopVirtualPRD}
\hrefCMSnoop {} {G.~L. Kane, E.~Kuflik, R.~Lu{ et~al.}, ``{Top channel for
  early SUSY discovery at the LHC}'',} \textit{ Phys. Rev. D} \textbf{ 84}
  (2011) 095004,
  \href{http://dx.doi.org/10.1103/PhysRevD.84.095004}{\doi{10.1103/PhysRevD.84.095004}},
\href{http://www.arXiv.org/abs/1101.1963}{\texttt{ arXiv:1101.1963}}.

\bibitem{T1tttt}
\hrefCMSnoop {} {{ LHC New Physics Working Group} Collaboration, ``{Simplified
  Models for LHC New Physics Searches}'',} (2011).
  \href{http://www.arXiv.org/abs/1105.2838}{\texttt{ arXiv:1105.2838}}.
Model of Section IV.E with topology (B+B).

\bibitem{wacker}
\hrefCMSnoop {} {R.~Essig, E.~Izaguirre, J.~Kaplan{ et~al.}, ``{Heavy flavor
  simplified models at the LHC}'',} \textit{ JHEP} \textbf{ 1201} (2012) 074,
  \href{http://dx.doi.org/10.1007/JHEP01(2012)074}{\doi{10.1007/JHEP01(2012)074}},
\href{http://www.arXiv.org/abs/1110.6443}{\texttt{ arXiv:1110.6443}}.

\bibitem{Kramer:2012bx}
\hrefCMSnoop {} {M.~Kramer, A.~Kulesza, R.~van~der Leeuw{ et~al.},
  ``{Supersymmetry production cross sections in pp collisions at sqrt{s} = 7
  TeV}'',} (2012).
\href{http://www.arXiv.org/abs/1206.2892}{\texttt{ arXiv:1206.2892}}.

\bibitem{Kulesza:2008jb}
\hrefCMSnoop {} {A.~Kulesza and L.~Motyka, ``{Threshold Resummation for
  Squark-Antisquark and Gluino-Pair Production at the LHC}'',} \textit{ Phys.
  Rev. Lett.} \textbf{ 102} (2009) 111802,
  \href{http://dx.doi.org/10.1103/PhysRevLett.102.111802}{\doi{10.1103/PhysRevLett.102.111802}},
\href{http://www.arXiv.org/abs/0807.2405}{\texttt{ arXiv:0807.2405}}.

\bibitem{Beenakker:2010nq}
\hrefCMSnoop {} {W.~Beenakker, S.~Brensing, M.~Kramer{ et~al.},
  ``{Supersymmetric top and bottom squark production at hadron colliders}'',}
  \textit{ JHEP} \textbf{ 1008} (2010) 098,
  \href{http://dx.doi.org/10.1007/JHEP08(2010)098}{\doi{10.1007/JHEP08(2010)098}},
\href{http://www.arXiv.org/abs/1006.4771}{\texttt{ arXiv:1006.4771}}.

\bibitem{ATLAS:2012ai}
\hrefCMSnoop {} {{ ATLAS} Collaboration, ``{Search for gluinos in events with
  two same-sign leptons, jets and missing transverse momentum with the ATLAS
  detector in pp collisions at sqrt(s) = 7 TeV}'',} \textit{ Phys. Rev. Lett.}
  \textbf{ 108} (2012) 241802,
\href{http://www.arXiv.org/abs/1203.5763}{\texttt{ arXiv:1203.5763}}.

\bibitem{ATLAS:2012ah}
\hrefCMSnoop {} {{ ATLAS} Collaboration, ``{Search for supersymmetry in pp
  collisions at sqrt(s) = 7 TeV in final states with missing transverse
  momentum and b-jets with the ATLAS detector}'',} \textit{ Phys. Rev.}
  \textbf{ D85} (2012) 112006,
\href{http://www.arXiv.org/abs/1203.6193}{\texttt{ arXiv:1203.6193}}.

\end{thebibliography}\endgroup

\cleardoublepage \appendix\section{The CMS Collaboration \label{app:collab}}\begin{sloppypar}\hyphenpenalty=5000\widowpenalty=500\clubpenalty=5000\textbf{Yerevan Physics Institute,  Yerevan,  Armenia}\\*[0pt]
S.~Chatrchyan, V.~Khachatryan, A.M.~Sirunyan, A.~Tumasyan
\vskip\cmsinstskip
\textbf{Institut f\"{u}r Hochenergiephysik der OeAW,  Wien,  Austria}\\*[0pt]
W.~Adam, T.~Bergauer, M.~Dragicevic, J.~Er\"{o}, C.~Fabjan\cmsAuthorMark{1}, M.~Friedl, R.~Fr\"{u}hwirth\cmsAuthorMark{1}, V.M.~Ghete, J.~Hammer, N.~H\"{o}rmann, J.~Hrubec, M.~Jeitler\cmsAuthorMark{1}, W.~Kiesenhofer, V.~Kn\"{u}nz, M.~Krammer\cmsAuthorMark{1}, D.~Liko, I.~Mikulec, M.~Pernicka$^{\textrm{\dag}}$, B.~Rahbaran, C.~Rohringer, H.~Rohringer, R.~Sch\"{o}fbeck, J.~Strauss, A.~Taurok, P.~Wagner, W.~Waltenberger, G.~Walzel, E.~Widl, C.-E.~Wulz\cmsAuthorMark{1}
% ...
\vskip\cmsinstskip
\dag:~Deceased\\
1:~~Also at Vienna University of Technology, Vienna, Austria\\
% ...

\end{sloppypar}
\end{document}